\documentclass[aps,showpacs,showkeys,preprintnumbers,amsmath,amssymb]
{revtex4}
\usepackage{graphicx}
\begin{document}

\title{
Bethe-Salpeter equation for doubly heavy baryons in the covariant
instantaneous approximation}

\author{ M.-H. Weng$^{1}$\footnote[1]{ \emph{E-mail address:}
mhweng@mail.bnu.edu.cn}, X.-H. Guo$^{1,2}$\footnote[2]{ \emph{E-mail
address:} xhguo@bnu.edu.cn}, A. W. Thomas$^{2}$\footnote[3]{
\emph{E-mail address:} anthony.thomas@adelaide.edu.au}}
\affiliation{$^{1}$ College of Nuclear Science and Technology,
Beijing Normal University, Beijing
100875, People's Republic of China\\
$^{2}$ CSSM, School of Chemistry and Physics, University of
Adelaide, Adelaide SA 5005, Australia }

%%%%%%%%%%%%%%%%%%%%%%%%%%%%%%%%%%%%%%%%%%%%%%%%%%%%%%

\begin{abstract}

In the heavy quark limit, a doubly heavy baryon is regarded as
composed of a heavy diquark and a light quark. We establish the
Bethe-Salpeter (BS) equations for the heavy diquarks and the doubly
heavy baryons, respectively, to leading order in a $1/m_{Q}$
expansion. The BS equations are solved numerically under the
covariant instantaneous approximation with the kernels containing
scalar confinement and one-gluon-exchange terms. The masses for the
heavy diquarks and the doubly heavy baryons are obtained and the
non-leptonic decay widths for the doubly heavy baryons emitting a
pseudo-scalar meson are calculated within the model.

\end{abstract}



%12.39.Hg    Heavy quark effective theory
%13.30.Eg    Hadronic decays
%14.20.Lq    Charmed baryons (|C|>0, B=0)
%14.20.Mr    Bottom baryons (|B|>0)
%11.10.St    Bound and unstable states; Bethe-Salpeter equations

\maketitle

\section*{I. Introduction}

The past few years have seen many important developments concerning
hadron colliders, especially the advent of the LHC. Recently, the
discovery  of and searches for double charm baryons have been
reported by various experimental collaborations
\cite{mat02,och05,chi06,aub06}. One is convinced that more and more
doubly heavy baryons will be observed in the near future.
Consequently, it is an urgent task for theorists to investigate the
properties of these states.

On the other hand, the existence of three valence quarks in a baryon
makes the theoretical study much more complicated than the case of
mesons. People have suggested the presence of diquark structure in a
baryon and studied the properties of heavy baryons in such a picture
\cite{ans87,guo96,ebe98,guo99,guo08}. No matter whether the diquark
is a real physical object or simply a theoretical approximation,
this picture reduces the three body system to a two body problem
which is much simpler for investigation.

In recent years, heavy quark effective theory (HQET) has been widely
used in the study of doubly heavy baryons
\cite{whi91,san95,guo98,ton00,dai00,ebe04,fly07,her08}. Two heavy
quarks are reasonably bound into a color-antitriplet heavy diquark
whose radius is much smaller than the typical scale
($1/\Lambda_{QCD}$) of the nonperturbative QCD interactions in the
heavy quark limit ($m_{Q}\gg \Lambda_{QCD}$, $m_{Q}$ denotes the
heavy quark mass). The leftover light quark involved in the baryon
moves in the color field induced by the heavy diquark. Unlike the
heavy quark and light quark system, the internal motion in the heavy
diquark can not be ignored even at leading order in the $1/m_{Q}$
expansion. This is because the relative momentum in the heavy
diquark is not simply ${\cal O}(\Lambda_{QCD})$, but
$\sim\alpha^{2}_{s}m_{Q}$ as calculated in the Coulomb potential
model \cite{san95,luc92}.

As a formally exact equation to describe the relativistic bound
system, the Bethe-Salpeter (BS) equation was initially formulated in
Minkowski space based on the relativistic quantum theory
\cite{sal51,nak69}. However, it is difficult to solve the BS
equation in Minkowski space due to its singular behavior. In order
to overcome this difficulty, with the so-called "Wick rotation", the
formalism for the BS equation in Euclidean space was developed and
investigated in detail \cite{wic54,nie96,efi03,rob07}. With the
perturbation theory integral representation, the BS equation was
solved in Minkowski space for the scalar and fermion systems
\cite{nak63,nak71,kus95,kus97,sau03,sau08}. Recently, based on the
Nakanishi integral representation of the BS amplitude and the
projection of the BS equation on the light-front plane, a new method
for solving the BS equation in Minkowski space was proposed and was
applied to study  the electromagnetic form factor
\cite{kar0601,kar0602,kar0603,car09,car1001,car1002}. In another
formalism to solve the BS equation, the covariant instantaneous
approximation is adopted in the kernel. In recent decades, this
formalism has been successfully used to investigate heavy mesons,
heavy baryons, and exotic states
\cite{jai,guo96,guo08,guo98,guo99,mir,guo07,guo10,jin92,dai93,dai94,cha05}.

Based on the diquark picture of the composition of the doubly heavy
baryon, we will establish the BS equations for both the heavy
diquarks and the doubly heavy baryons in the leading order of a
$1/m_{Q}$ expansion. Motivated by the potential model, the kernel
for the BS equation is assumed to be composed of the scalar
confinement and one-gluon-exchange terms \cite{eic78}. We will solve
the BS equations numerically under the covariant instantaneous
approximation \cite{jin92,dai93,dai94,cha05}. Since the heavy
diquark is not really a point object, a few form factors for the
effective vertex of the heavy diquark coupling to the gluon are
introduced to reflect the inner structure of the heavy diquark.
These form factors will be expressed in terms of the BS wave
functions obtained for the heavy diquarks. Finally, we will
calculate the non-leptonic decay widths for the doubly heavy baryons
emitting a pseudo-scalar meson in the BS formalism.

The remainder of this paper is organized as follows. In Sec. II, we
establish the BS equation for the heavy diquarks in the leading
order of $1/m_{Q}$ expansion. We also give the normalization
conditions for the BS wave functions for the heavy diquarks in this
section. In Sec. III, the form factors for the effective vertex of
the heavy diquark coupling to the gluon are derived from the BS
wave functions obtained for the heavy diquarks. In Sec. IV, we establish the
BS equation for the doubly heavy baryons at leading order in
the $1/m_{Q}$ expansion. The normalization conditions for the BS wave
functions for the doubly heavy baryons are also given in this
section. In Sec. V, the non-leptonic decay widths for the doubly
heavy baryons emitting a pseudo-scalar meson are calculated in the
BS formalism. Sec. VI is reserved for our summary and some
discussions.

\section*{II. BS equation for heavy diquarks}

In general, the parity of a ground state baryon is positive. Since
the parity of quark is supposed to be positive, the parity of the
diquark involved in a ground state baryon should be positive. Due to
the Pauli principle, two quarks with the same flavor can only
constitute an axial-vector diquark. On the other hand, two quarks
with different flavors can constitute either a scalar diquark or an
axial-vector diquark. It can be easily shown that a heavy diquark
which is in the ground state can not be a tensor diquark.

Suppose two heavy quarks $Q_{1}$ and $Q_{2}$ (with masses
$m_{Q_{1}}$ and $m_{Q_{2}}$ respectively) compose a ground state
heavy diquark. Define two ratios
$\lambda_{1}=m_{Q_{1}}/(m_{Q_{1}}+m_{Q_{2}})$ and
$\lambda_{2}=m_{Q_{2}}/(m_{Q_{1}}+m_{Q_{2}})$. The BS wave function
for the heavy diquark is defined as follows:
\begin{eqnarray}\label{eq:1-2}
  \chi_{P_{D}}(x_{1},x_{2})_{\alpha\beta}&=&\varepsilon^{ijk}\langle0
  |T\psi_{1}(x_{1})^{i}_{\alpha}\psi_{2}(x_{2})^{j}_{\beta}|P_{D},k\rangle\nonumber\\
  &=&e^{-iP_{D}X}\int\frac{{\rm
  d}^{4}p}{(2\pi)^{4}}\chi_{P_{D}}(p)_{\alpha\beta}e^{-ipx},\\
\bar{\chi}_{P_{D}}(x_{2},x_{1})_{\beta\alpha}&=&\varepsilon^{ijk}\langle
P_{D},k
  |T\psi^{\ast}_{2}(x_{2})^{j}_{\beta}\psi^{\ast}_{1}(x_{1})^{i}_{\alpha}|0\rangle\nonumber\\
  &=&e^{iP_{D}X}\int\frac{{\rm
  d}^{4}p}{(2\pi)^{4}}{\bar\chi}_{P_{D}}(p)_{\beta\alpha}e^{ipx},
\end{eqnarray}
where $\psi_{1}$ and $\psi_{2}$ stand for the field operators of the
heavy quarks $Q_{1}$ and $Q_{2}$, respectively, $i,~j,~k$ represent
the color indices, $\alpha$ and $\beta$ represent the spin indices,
$X\equiv\lambda_{1}x_{1}+\lambda_{2}x_{2}$ is the coordinate of the
heavy diquark mass center, $x\equiv x_{1}-x_{2}$ is the relative
coordinate of the two heavy quarks, $P_{D}$ is the momentum of the
heavy diquark, and $p$ is the relative momentum between the two
heavy quarks.

The BS equation for the heavy diquark can be written in the
following form (details can be found in Ref. \cite{guo08}):
\begin{eqnarray}\label{eq:3}
  \chi_{P_{D}}(p)&=&S(p_{1})\otimes S(p_{2})
  \int\frac{{\rm d}^{4}p^{\prime}}{(2\pi)^{4}}
  [\gamma^{\mu}\otimes\gamma_{\mu}K^{(1g)}(p-p^{\prime})
  +I\otimes IK^{(cf)}(p-p^{\prime})]
  \chi_{P_{D}}(p^{\prime}),
\end{eqnarray}
where $p_{1}=\lambda_{1}P_{D}+p$ and $p_{2}=\lambda_{2}P_{D}-p$ are
the momenta of heavy quarks $Q_{1}$ and $Q_{2}$, respectively,
$S(p_{1})$ and $S(p_{2})$ are the propagators of heavy quarks
$Q_{1}$ and $Q_{2}$, respectively. $K^{(1g)}$ and $K^{(cf)}$ are the
one-gluon-exchange and scalar confinement terms of the kernel for
the BS equation given by (after imposing the covariant
instantaneous approximation \cite{jin92,dai93,dai94,cha05}):
\begin{equation}\label{eq:4}
  K^{(1g)}(p_{t}-p^{\prime}_{t})=-\frac{8i\pi}{3}\frac{\alpha_{s}}{(p_{t}-p_{t}^{\prime})^{2}-\mu^{2}},
\end{equation}
and
\begin{equation}\label{eq:5}
  K^{(cf)}(p_{t}-p_{t}^{\prime})=\frac{4i\pi\kappa}{[-(p_{t}-p_{t}^{\prime})^2+\mu^{2}]^{2}}-(2\pi)^{3}\delta^{3}(p_{t}-p_{t}^{\prime})
  \int\frac{{\rm
  d}^{3}k_{t}}{(2\pi)^{3}}\frac{4i\pi\kappa}{[-(p_{t}-k_{t})^2+\mu^{2}]^{2}},
\end{equation}
where $\alpha_{s}$ and $\kappa$ are the coupling parameters related
to one-gluon-exchange and scalar confinement terms, respectively,
$p_{t}$ is the transverse projection of the relative momentum ($p$)
along the heavy diquark momentum ($P_{D}$) (see the definition below
Eq. (\ref{eq:8})), the second term of $K^{(cf)}$ is introduced to
remove the infrared singularity near the point
$p^{\prime}_{t}=p_{t}$, and the small parameter $\mu$ is introduced
to avoid the divergence in the numerical calculations. This kernel
is motivated by the potential model which has been successfully
applied in mesons \cite{eic78}. Furthermore, we assume that the
kernel of the heavy diquark is related to the meson by the one-half
rule \cite{cor96,ebe97}.

Eq. (\ref{eq:3}) can be written in a more usual matrix form as
\cite{guo08}
\begin{eqnarray}\label{eq:6}
  {\tilde \chi}_{P_{D}}^{T}(p)&=&S(p_{2})\int\frac{{\rm
  d}^{4}p^{\prime}}{(2\pi)^{4}}[-\gamma^{\mu}{\tilde \chi}_{P_{D}}^{T}(p^{\prime})
  \gamma_{\mu}K^{(1g)}(p_{t}-p_{t}^{\prime})
  +{\tilde
  \chi}_{P_{D}}^{T}(p^{\prime})K^{(cf)}(p_{t}-p_{t}^{\prime})]S(-p_{1}),
\end{eqnarray}
where ${\tilde
  \chi}_{P_{D}}(p)={\cal C}\chi_{P_{D}}(p)$ (${\cal C}$ is the charge conjugation
matrix) and the superscript $T$ represents the transpose of the spinor
indices.

In the leading order of a $1/m_{Q}$ expansion, the heavy quark
propagators ($S(p_{1})$ and $S(p_{2})$) can be written as
\begin{eqnarray}\label{eq:7}
  S(p_{1})&=&i\frac{m_{Q_{1}}(\rlap/v_{D}+1)}{2\omega_{Q_{1}}(\lambda_{1}m_{D}+p_{l}-\omega_{Q_{1}}+i\varepsilon)},
\end{eqnarray}
and
\begin{eqnarray}\label{eq:8}
  S(p_{2})&=&-i\frac{m_{Q_{2}}(\rlap/v_{D}+1)}{2\omega_{Q_{2}}(-\lambda_{2}m_{D}+p_{l}+\omega_{Q_{2}}-i\varepsilon)},
\end{eqnarray}
where $m_{D}$ and $v_{D}$ are the mass and velocity of the heavy
diquark, respectively, $ p_{l}=p\cdot v_{D}$ and
$p^{\mu}_{t}=p^{\mu}-p_{l}v^{\mu}_{D}$ are the longitudinal and
transverse projections of the relative momentum ($p$) along the
heavy diquark momentum
 ($P_{D}$), respectively, the energy $\omega_{Q_{1(2)}}=\sqrt{m_{Q_{1(2)}}^{2}-p_{t}^{2}}$,
and $\varepsilon$ is the infinitesimal. As studied in Refs.
\cite{whi91,san95}, $|p_{t}|$ is ${\cal O}(\alpha_{s}^{2}m_{Q})$.
Consequently, $|p_{t}|/m_{Q}$ terms should not be neglected in calculations
carried out to leading order in the $1/m_{Q}$ expansion.

Substituting Eqs. (\ref{eq:7}) (\ref{eq:8}) into Eq. (\ref{eq:6}),
one finds the following two constraints for the BS wave function
for the heavy diquarks:
\begin{eqnarray}\label{eq:9-10}
\rlap/v_{D}{\tilde \chi}_{P_{D}}^{T}(p)&=&{\tilde
\chi}_{P_{D}}^{T}(p),\\
{\tilde \chi}_{P_{D}}^{T}(p)\rlap/v_{D}&=&-{\tilde
\chi}_{P_{D}}^{T}(p).
\end{eqnarray}

Then, taking these constraints into account
in the BS equation for the positive parity and zero
angular momentum ground state of the heavy diquark system, the BS
wave functions for the scalar and the axial-vector heavy diquarks
can be parametrized in the following forms, respectively:
\begin{eqnarray}\label{eq:11}
  {\tilde \chi}_{P_{D}}^{T}(p)&=&(\rlap/v_{D}+1)\gamma^{5}f_{1},
\end{eqnarray}
and
\begin{eqnarray}\label{eq:12}
  {\tilde \chi}_{P_{D}}^{(r)T}(p)&=&(\rlap/v_{D}+1)\rlap/\xi^{(r)} f_{2},
\end{eqnarray}
where $\xi^{(r)}_{\mu}$ is the $r$-th polarization vector of the
axial-vector heavy diquark, $f_{1}$ and $f_{2}$ are the
Lorentz-scalar functions of $p_{t}^{2}$, $p_{l}$, and
$P_{D}^{2}=m_{D}^{2}$.

After some algebra, we find that the BS scalar wave functions for
both the scalar heavy diquark ($f_{1}$) and the axial-vector heavy
diquark ($f_{2}$) satisfy the same integral equation as follows:
\begin{eqnarray}\label{eq:13}
  {\tilde f} (p_{t})=\frac{m_{Q_{1}}m_{Q_{2}}}{\omega_{Q_{1}}\omega_{Q_{2}}
  (-m_{D}+\omega_{Q_{1}}+\omega_{Q_{2}})}\int\frac{{\rm
  d}^{3}p^{\prime}_{t}}{(2\pi)^{3}}[V^{(1g)}(p_{t}-p^{\prime}_{t})
  +V^{(cf)}(p_{t}-p^{\prime}_{t})]{\tilde f}(p^{\prime}_{t}),
\end{eqnarray}
where we define ${\tilde f} (p_{t})\equiv\int\frac{{\rm
d}p_{l}}{2\pi}f_{1(2)}(p)$, $V^{(1g)}\equiv-iK^{(1g)}$, and
$V^{(cf)}\equiv-iK^{(cf)}$.

In general, the normalization condition for the heavy diquark can be
written as (after imposing the covariant instantaneous approximation
on the kernel) \cite{guo08}
\begin{eqnarray}\label{eq:14}
\frac{i}{36}\delta^{i_{1}i_{2}}_{j_{1}j_{2}}\int\frac{{\rm
  d}^{4}p{\rm
  d}^{4}p^{\prime}}{(2\pi)^{8}}\bar{\chi}_{P_{D}}(p)
  \frac{\partial}{\partial P_{D}^{0}}\Bigl[I_{P_{D}}(p,p^{\prime})
  \Bigr]^{i_{1}i_{2}j_{2}j_{1}}\chi_{P_{D}}(p^{\prime})=1,
\end{eqnarray}
where $i_{1(2)}$ and $j_{1(2)}$ represent the color indices of the
heavy quarks, $\delta^{i_{1}i_{2}}_{j_{1}j_{2}}
=\delta^{i_{1}}_{j_{1}}\delta^{i_{2}}_{j_{2}}-
\delta^{i_{1}}_{j_{2}}\delta^{i_{2}}_{j_{1}}$, %$K(p,p^{\prime})$ is
%the kernel for the BS equation,
and $I_{P_{D}}^{i_{1}i_{2}j_{2}j_{1}}(p,p^{\prime})$ stands for the
inverse of the four point function,
\begin{eqnarray}\label{eq:15}
 I_{P_{D}}^{i_{1}i_{2}j_{2}j_{1}}(p,p^{\prime})
=\delta^{i_{1}j_{1}}\delta^{i_{2}j_{2}}(2\pi)^{4}\delta^{4}(p-p^{\prime})
[S(p_{1})\gamma_{0}]^{-1}[S(p_{2})\gamma_{0}]^{-1}.
\end{eqnarray}

Now, it is straightforward to obtain the normalization condition for
the BS wave function for the heavy diquark as the following:
\begin{eqnarray}\label{eq:16}
  -\frac{1}{6}\int\frac{{\rm
  d}^{4}p}{(2\pi)^{4}}\Bigl\{{\rm
  Tr}[-\lambda_{1}S(-p_{1}){\tilde{\tilde
  \chi}}_{P_{D}}^{(c)}(p_{t})S(p_{2}){\tilde{\tilde
  \chi}}_{P_{D}}(p_{t})S(-p_{1})\rlap/\varepsilon]\nonumber\\
  +{\rm
  Tr}[\lambda_{2}S(-p_{1}){\tilde{\tilde
  \chi}}_{P_{D}}^{(c)}(p_{t})S(p_{2})\rlap/\varepsilon S(p_{2}){\tilde{\tilde
  \chi}}_{P_{D}}(p_{t})]\Bigl\}=1,
\end{eqnarray}
where $\varepsilon=(1,{\vec 0})$, ${\tilde{\tilde
  \chi}}_{P_{D}}(p_{t})$ and ${\tilde{\tilde
  \chi}}_{P_{D}}^{(c)}(p_{t})$ are the transverse projections of the BS
  wave function given by
\begin{eqnarray}\label{eq:17}
{\tilde{\tilde
  \chi}}_{P_{D}}(p_{t})&=&-i S(p_{2})^{-1}{\tilde
  \chi}_{P_{D}}^{T}(p)S(-p_{1})^{-1},
%{\tilde \chi}_{P}^{T}(p)&=&i S(p_{2}){\tilde{\tilde
%  \chi}}_{P}(p_{t})S(-p_{1}),\\
\end{eqnarray}
and
\begin{eqnarray}\label{eq:18}
{\tilde{\tilde
  \chi}}_{P_{D}}^{(c)}(p_{t})&=&{\cal C}{\tilde{\tilde
  \chi}}_{-P_{D}}^{T}(-p_{t}){\cal C}^{-1},
\end{eqnarray}
respectively.

For the scalar heavy diquark, the transverse projections of the BS
wave function (${\tilde{\tilde
  \chi}}_{P_{D}}(p_{t})$ and ${\tilde{\tilde
  \chi}}_{P_{D}}^{(c)}(p_{t})$) can be obtained from Eqs. (\ref{eq:17}) (\ref{eq:18})
  as follows respectively:
\begin{eqnarray}\label{eq:19}
 {\tilde{\tilde
  \chi}}_{P_{D}}(p_{t})={\tilde f_{s_{1}}}(p_{t})\gamma^{5}+{\tilde
  f_{s_{2}}}(p_{t})\rlap/v_{D}\gamma^{5},
\end{eqnarray}
and
\begin{eqnarray}\label{eq:20}
{\tilde{\tilde
  \chi}}_{P_{D}}^{(c)}(p_{t})={\tilde f_{s_{1}}}(p_{t})\gamma^{5}-{\tilde
  f_{s_{2}}}(p_{t})\rlap/v_{D}\gamma^{5},
\end{eqnarray}
where
\begin{eqnarray}\label{eq:21}
  {\tilde f_{s_{1}}}(p_{t})&=&\int\frac{{\rm
  d}^{3}p^{\prime}_{t}}{(2\pi)^{3}}[V^{(1g)}(p_{t}-p^{\prime}_{t})
  +4V^{(cf)}(p_{t}-p^{\prime}_{t})]{\tilde f}(p^{\prime}_{t}),
  \end{eqnarray}
  and
  \begin{eqnarray}\label{eq:22}
  {\tilde f_{s_{2}}}(p_{t})&=&\int\frac{{\rm
  d}^{3}p^{\prime}_{t}}{(2\pi)^{3}}[V^{(1g)}(p_{t}-p^{\prime}_{t})
  -2V^{(cf)}(p_{t}-p^{\prime}_{t})]{\tilde f}(p^{\prime}_{t}).
\end{eqnarray}

After substituting Eqs. (\ref{eq:19}) (\ref{eq:20}) into Eq.
(\ref{eq:16}), carrying out the trace calculation, and integrating
out the longitudinal momentum $p_{l}$, we obtain the normalization
condition for the BS wave function for the scalar heavy diquark as
the following:
\begin{eqnarray}\label{eq:23}
  \int\frac{{\rm
  d}^{3}p_{t}}{(2\pi)^{3}}\frac{E_{D}m_{Q_{1}}m_{Q_{2}}(\lambda_{1}m_{Q_{1}}\omega_{Q_{2}}
  +\lambda_{2}m_{Q_{2}}\omega_{Q_{1}})}{3m_{D}\omega^{2}_{Q_{1}}\omega^{2}_{Q_{2}}
  (-m_{D}+\omega_{Q_{1}}+\omega_{Q_{2}})^{2}}[{\tilde f}_{s_{1}}(p_{t})+{\tilde f}_{s_{2}}(p_{t})]^{2}=1,
\end{eqnarray}
where $E_{D}=P_{D}\cdot \varepsilon$.

For the axial-vector heavy diquark, the transverse projections of
the BS wave function (${\tilde{\tilde
  \chi}}^{(r)}_{P_{D}}(p_{t})$ and ${\tilde{\tilde
  \chi}}^{(r)(c)}_{P_{D}}(p_{t})$) can be obtained from Eqs. (\ref{eq:17}) (\ref{eq:18})
  as follows respectively:
\begin{eqnarray}\label{eq:24}
 {\tilde{\tilde
  \chi}}^{(r)}_{P_{D}}(p_{t})={\tilde f_{v_{1}}}(p_{t})\rlap/\xi^{(r)}+{\tilde
  f_{v_{2}}}(p_{t})\rlap/v_{D}\rlap/\xi^{(r)},
\end{eqnarray}
and
\begin{eqnarray}\label{eq:25}
 {\tilde{\tilde
  \chi}}^{(r)(c)}_{P_{D}}(p_{t})=-{\tilde f_{v_{1}}}(p_{t})\rlap/\xi^{(r)}+{\tilde
  f_{v_{2}}}(p_{t})\rlap/v_{D}\rlap/\xi^{(r)},
\end{eqnarray}
 where
\begin{eqnarray}\label{eq:26}
  {\tilde f_{v_{1}}}(p_{t})&=&\int\frac{{\rm
  d}^{3}p^{\prime}_{t}}{(2\pi)^{3}}[V^{(1g)}(p_{t}-p^{\prime}_{t})
  +2V^{(cf)}(p_{t}-p^{\prime}_{t})]{\tilde f}(p^{\prime}_{t}),
  \end{eqnarray}
  and
  \begin{eqnarray}\label{eq:27}
  {\tilde f_{v_{2}}}(p_{t})&=&\int\frac{{\rm
  d}^{3}p^{\prime}_{t}}{(2\pi)^{3}}V^{(1g)}(p_{t}-p^{\prime}_{t}){\tilde f}
  (p^{\prime}_{t}).
\end{eqnarray}

Analogously, the normalization condition for the BS wave function
for the axial-vector heavy diquark is given by
\begin{eqnarray}\label{eq:28}
  \int\frac{{\rm
  d}^{3}p_{t}}{(2\pi)^{3}}\frac{E_{D}m_{Q_{1}}m_{Q_{2}}(\lambda_{1}m_{Q_{1}}\omega_{Q_{2}}
  +\lambda_{2}m_{Q_{2}}\omega_{Q_{1}})}{3m_{D}\omega^{2}_{Q_{1}}\omega^{2}_{Q_{2}}
  (-m_{D}+\omega_{Q_{1}}+\omega_{Q_{2}})^{2}}
  [{\tilde f}_{v_{1}}(p_{t})+{\tilde f}_{v_{2}}(p_{t})]^{2}=1.
\end{eqnarray}

In the numerical calculations, we take the constituent masses of the
heavy quarks to be $m_{b}=4.88~GeV$ and $m_{c}=1.486~GeV$ which were
obtained by fitting the real spectra of charmonium and bottomonium
in Ref. \cite{ger00}. The parameters in the kernel $\alpha_{s}=0.4$
and $\kappa=0.18$ are determined by fitting the experimental data
for heavy meson spectra \cite{eic78}. In order to solve the integral
equation (\ref{eq:13}), we discretize the integration region into
$n$ pieces (with $n$ sufficiently large). Then Eq.~(\ref{eq:13})
becomes an eigenvalue equation for the $n$ dimensional vector
$\tilde{f}$. After solving the eigenvalue equation, the heavy
diquark masses are obtained and are displayed in Table~\ref{tab:1}.
We find that the heavy diquark masses are independent of the heavy
diquark spin and only determined by the flavors of the constituent
heavy quarks. In Fig.~1, the normalized BS scalar wave functions for
the heavy diquarks are shown. It can be seen that the amplitudes of
the BS scalar wave functions do not distinguish the different spins
of the heavy diquarks. As discussed in Refs.~\cite{whi91,san95},
unlike the heavy quark and light quark system, only the spin
symmetry survives (when the diquark mass is below $\sim10~GeV$) in
the leading order $1/m_{Q}$ expansion for the heavy diquark system.
Our results are consistent with this statement.

\begin{table}[hbtp]
\caption{\label{tab:1}Values of the heavy diquark masses used here.}
\begin{ruledtabular}
\begin{tabular}{ccccccc}
             $m_{Q_{1}}~(GeV)$ &4.88  &4.88  &1.486\\
             $m_{Q_{2}}~(GeV)$ &4.88  &1.486 &1.486\\
\hline       $m_{D}~(GeV)$     &9.80  &6.55  &3.23\\

\end{tabular}
\end{ruledtabular}
\end{table}

\begin{figure}[hbtp]\label{fig1}
\begin{center}
\includegraphics[scale=0.75]{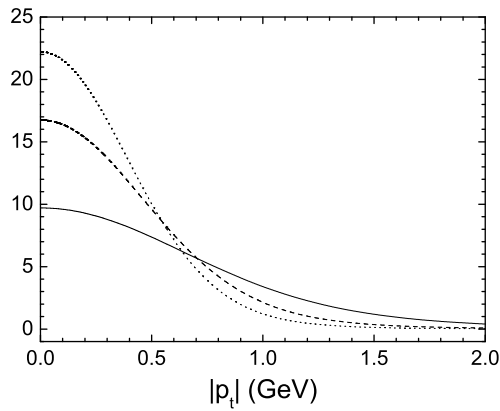}
\caption{The normalized BS scalar wave functions (${\tilde
f}(p_{t})$) for the heavy diquarks. The solid, dashed, and dotted
lines are for the heavy diquarks composed of double $b$ quarks, $b$
and $c$ quarks, and double $c$ quarks, respectively.}
\end{center}
\end{figure}

\section*{III. Form factor of heavy diquark coupling to gluon}

Since the heavy diquark is not really a point object and its radius
is enhanced by ${\rm ln}^{2}m_{Q}$ with respect to $1/m_{Q}$, we
introduce a few form factors for the effective vertex of the heavy
diquark coupling to gluon to reflect the inner structure of the
heavy diquark.

\begin{figure}[hbtp]\label{fig2}
\begin{center}
\includegraphics[scale=0.8]{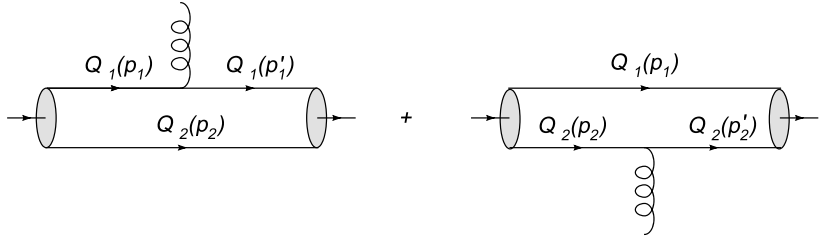}
\caption{The schematic diagram for the heavy diquark coupling to
gluon. $Q_{1}(p_{1}^{(\prime)})$ and $Q_{2}(p_{2}^{(\prime)})$ stand
for the heavy quarks $Q_{1}$ and $Q_{2}$ with momenta
$p_{1}^{(\prime)}$ and $p_{2}^{(\prime)}$, respectively.}
\end{center}
\end{figure}

The effective current for the scalar heavy diquark coupling to a gluon
is given as follows~\cite{ans87}:
\begin{eqnarray}\label{eq:29}
  J^{\mu}&=&
  ig_{s}\frac{\lambda^{a}}{2}(P_{D_{f}}^{\mu}+P_{D_{i}}^{\mu})F_{s}(Q^2),
\end{eqnarray}
where $g_{s}$ is the coupling constant of the strong interaction,
$\lambda^{a}$ ($a=1,~2,\cdot\cdot\cdot,~8$) denote the Gell-Mann
color matrices, $F_{s}(Q^{2})$ is the form factor for the effective
vertex, $P_{D_{i}}^{\mu}$ and $P_{D_{f}}^{\mu}$ are the momenta of
the initial and final heavy diquarks, respectively, and $Q^2$ is the
square of the momentum transfer.

On the other hand, the effective current for the scalar heavy diquark
coupling to a gluon can be written as the following in the BS equation
formalism (The Feynmann diagram for the vertex is shown in Fig. 2):
\begin{eqnarray}\label{eq:30}
  J^{\mu}
  &=&ig_{s}\frac{\lambda^{a}}{2}(M^{\mu}_{1}+M^{\mu}_{2}),
\end{eqnarray}
where
\begin{eqnarray}\label{eq:31}
  M^{\mu}_{1}&=&-\frac{1}{6}(2\pi)^{4}\delta^{4}(P_{D_{i}}-P_{D_{f}}-k)\int\frac{{\rm
d}^{4}p}{(2\pi)^{4}}\int\frac{{\rm
d}^{4}p^{\prime}}{(2\pi)^{4}}(2\pi)^{4}\delta^{4}(p_{2}-p_{2}^{\prime})\nonumber\\
&&\times{\rm Tr}[S(-p_{1}){\tilde{\tilde
  \chi}}_{P_{D_{f}}}^{(c)}(p_{t})S(p_{2}){\tilde{\tilde
  \chi}}_{P_{D_{i}}}(p_{t}^{\prime})S(-p_{1}^{\prime})v^{\mu}],
\end{eqnarray}
and
\begin{eqnarray}\label{eq:32}
M^{\mu}_{2}&=&-\frac{1}{6}(2\pi)^{4}\delta^{4}(P_{D_{i}}-P_{D_{f}}-k)\int\frac{{\rm
d}^{4}p}{(2\pi)^{4}}\int\frac{{\rm
d}^{4}p^{\prime\prime}}{(2\pi)^{4}}(2\pi)^{4}\delta^{4}(p_{1}-p_{1}^{\prime\prime})\nonumber\\
&&\times{\rm Tr}[S(-p_{1}){\tilde{\tilde
  \chi}}_{P_{D_{f}}}^{(c)}(p_{t})S(p_{2})v^{\mu}S(p_{2}^{\prime\prime}){\tilde{\tilde
  \chi}}_{P_{D_{i}}}(p_{t}^{\prime\prime})],
\end{eqnarray}
where $k$ denotes the momentum carried by the gluon, $p$ and
$p^{\prime(\prime\prime)}$ in this section denote the relative
momenta of final and initial heavy diquarks, respectively.

Then, comparing Eq.~(\ref{eq:29}) with Eq.~(\ref{eq:30}), one can
get the form factor ($F_{s}(Q^{2})$) in the leading order of a
$1/m_{Q}$ expansion as follows:
\begin{eqnarray}\label{eq:33}
  F_{s}(Q^{2})&=&-\frac{(2\pi)^{4}}{3m_{D}}\int\frac{{\rm
d}^{3}p_{t}}{(2\pi)^{3}}\Bigl\{\frac{m^{2}_{Q_{1}}m_{Q_{2}} [{\tilde
f}_{s1}(p_{t})+{\tilde f}_{s2}(p_{t})][{\tilde
f}_{s1}(p^{\prime}_{t})+{\tilde f}_{s2}(p^{\prime}_{t})]}
{\omega_{Q_{1}}\omega_{Q_{2}}\omega^{\prime}_{Q_{1}}(m_{D}-\omega_{Q_{1}}-\omega_{Q_{2}})
(m_{D}-\omega_{Q_{2}}{\tilde \omega}-|p_{t}|\sqrt{{\tilde
\omega}^{2}-1}cos\theta-\omega^{\prime}_{Q_{1}})}\nonumber\\
&&+\frac{m_{Q_{1}}m^{2}_{Q_{2}} [{\tilde f}_{s1}(p_{t})+{\tilde
f}_{s2}(p_{t})][{\tilde f}_{s1}(p^{\prime\prime}_{t})+{\tilde
f}_{s2}(p^{\prime\prime}_{t})]}
{\omega_{Q_{1}}\omega_{Q_{2}}\omega^{\prime\prime}_{Q_{2}}(m_{D}-\omega_{Q_{1}}-\omega_{Q_{2}})
(m_{D}-\omega_{Q_{1}}{\tilde \omega}+|p_{t}|\sqrt{{\tilde
\omega}^{2}-1}cos\theta-\omega^{\prime\prime}_{Q_{2}})}\Bigr\},
\end{eqnarray}
where the velocity transfer ${\tilde \omega}=v_{D_{i}}\cdot
v_{D_{f}}$ ($v_{D_{i}}$ and $v_{D_{f}}$ are the velocities of the
initial and final heavy diquarks respectively),
$p^{\prime(\prime\prime)}_{l}=p^{\prime(\prime\prime)}\cdot
v_{D_{i}}$ and
$p^{\prime(\prime\prime)}_{t}=p^{\prime(\prime\prime)}-p^{\prime(\prime\prime)}_{l}v_{D_{i}}$
 are the longitudinal and transverse projections of the initial
 heavy diquark
 relative momenta ($p^{\prime(\prime\prime)}$) along their momenta ($P_{D_{i}}$) respectively,
 $\omega^{\prime}_{Q_{1}}=\sqrt{m_{Q_{1}}^{2}-p_{t}^{\prime2}}$,
 $\omega^{\prime\prime}_{Q_{2}}=\sqrt{m_{Q_{2}}^{2}-p_{t}^{\prime\prime2}}$,
 and $\theta$ is the angle between $p_{t}$
 and $P_{D_{it}}^{\mu}(=P_{D_{i}}^{\mu}-v_{D_{f}}^{\mu}P_{D_{i}}\cdot v_{D_{f}})$.

Now, let us turn to the effective current of the axial-vector heavy
diquark coupling to gluon \cite{ans87}:
\begin{eqnarray}\label{eq:34}
   J^{\alpha\mu\beta}=ig_{s}\frac{\lambda^{a}}{2}\Bigl[g^{\alpha\beta}(P_{D_{f}}^{\mu}+P_{D_{i}}^{\mu})F_{v_{1}}(Q^{2})
   -(P_{D_{f}}^{\beta}g^{\mu\alpha}+P_{D_{i}}^{\alpha}g^{\mu\beta})F_{v_{2}}(Q^{2})
   +P_{D_{i}}^{\alpha}P_{D_{f}}^{\beta}(P_{D_{f}}^{\mu}+P_{D_{i}}^{\mu})F_{v_{3}}(Q^{2})\Bigr],
\end{eqnarray}
where $F_{v_{1}}(Q^{2})$, $F_{v_{2}}(Q^{2})$, and $F_{v_{3}}(Q^{2})$
are the form factors for the effective vertex.

Analogously, the effective current of the axial-vector heavy diquark
coupling to the gluon can be written as follows in the BS equation
formalism:
\begin{eqnarray}\label{eq:35}
  J^{\alpha\mu\beta}
  &=&ig_{s}\frac{\lambda^{a}}{2}(M^{\alpha\mu\beta}_{1}+M^{\alpha\mu\beta}_{2}),
\end{eqnarray}
where
\begin{eqnarray}\label{eq:36}
  M^{\alpha\mu\beta}_{1}&=&-\frac{1}{6}(2\pi)^{4}\delta^{4}(P_{D_{i}}-P_{D_{f}}-k)\int\frac{{\rm
d}^{4}p}{(2\pi)^{4}}\int\frac{{\rm
d}^{4}p^{\prime}}{(2\pi)^{4}}(2\pi)^{4}\delta^{4}(p_{2}-p_{2}^{\prime})\nonumber\\
&&\times{\rm Tr}[S(-p_{1}){\tilde{\tilde
  \chi}}_{P_{D_{f}}}^{(c)\alpha}(p_{t})S(p_{2}){\tilde{\tilde
  \chi}}^{\beta}_{P_{D_{i}}}(p_{t}^{\prime})S(-p_{1}^{\prime})v^{\mu}],
  \end{eqnarray}
  and
  \begin{eqnarray}\label{eq:37}
M^{\alpha\mu\beta}_{2}&=&-\frac{1}{6}(2\pi)^{4}\delta^{4}(P_{D_{i}}-P_{D_{f}}-k)\int\frac{{\rm
d}^{4}p}{(2\pi)^{4}}\int\frac{{\rm
d}^{4}p^{\prime\prime}}{(2\pi)^{4}}(2\pi)^{4}\delta^{4}(p_{1}-p_{1}^{\prime\prime})\nonumber\\
&&\times {\rm Tr}[S(-p_{1}){\tilde{\tilde
  \chi}}_{P_{D_{f}}}^{(c)\alpha}(p_{t})S(p_{2})v^{\mu}S(p_{2}^{\prime\prime}){\tilde{\tilde
  \chi}}^{\beta}_{P_{D_{i}}}(p_{t}^{\prime\prime})].
\end{eqnarray}

As discussed in Ref.~\cite{ans87}, the contribution of
the $F_{v_{3}}(Q^{2})$ term is suppressed at small and intermediate
 momentum transfer, $Q^{2}$, since such a term is multiplied by high powers of momenta.
Consequently, $F_{v_{3}}(Q^{2})$ is ignored in our calculation.
Comparing Eq.~(\ref{eq:34}) with Eq. (\ref{eq:35}), we derive the
other two form factors ($F_{v_{1}}(Q^{2})$ and $F_{v_{2}}(Q^{2})$)
in the leading order of a $1/m_{Q}$ expansion as follows:
\begin{eqnarray}\label{eq:38}
  F_{v_{1}}(Q^{2})
  &=&-\frac{(2\pi)^{4}}{12m_{D}}\int\frac{{\rm
d}^{3}p_{t}}{(2\pi)^{3}}\Bigl\{\frac{m^{2}_{Q_{1}}m_{Q_{2}} [{\tilde
f}_{v_{1}}(p_{t})+{\tilde f}_{v_{2}}(p_{t})][{\tilde
f}_{v_{1}}(p^{\prime}_{t})+{\tilde f}_{v_{2}}(p^{\prime}_{t})]}
{\omega_{Q_{1}}\omega_{Q_{2}}\omega^{\prime}_{Q_{1}}
(m_{D}-\omega_{Q_{1}}-\omega_{Q_{2}}) (m_{D}-\omega_{Q_{2}}{\tilde
\omega}-|p_{t}|\sqrt{{\tilde
\omega}^{2}-1}cos\theta-\omega^{\prime}_{Q_{1}})}\nonumber\\
&&+\frac{m_{Q_{1}}m^{2}_{Q_{2}} [{\tilde f}_{v_{1}}(p_{t})+{\tilde
f}_{v_{2}}(p_{t})][{\tilde f}_{v_{1}}(p^{\prime\prime}_{t})+{\tilde
f}_{v_{2}}(p^{\prime\prime}_{t})]}
{\omega_{Q_{1}}\omega_{Q_{2}}\omega^{\prime\prime}_{Q_{2}}
(m_{D}-\omega_{Q_{1}}-\omega_{Q_{2}}) (m_{D}-\omega_{Q_{1}}{\tilde
\omega}+|p_{t}|\sqrt{{\tilde
\omega}^{2}-1}cos\theta-\omega^{\prime\prime}_{Q_{2}})}\Bigr\},\nonumber\\
F_{v_{2}}(Q^{2})&=&0.
\end{eqnarray}

It can be seen that the form factors for both the effective vertex
of the scalar heavy diquark coupling to the gluon and the effective
vertex of the axial-vector heavy diquark coupling to the gluon are
equal to each other in the leading order of a $1/m_{Q}$ expansion.
So, we redefine the form factors as $F(Q^{2}) \equiv F_{s}(Q^{2}) =
F_{v_{1}}(Q^{2})$ for convenience.

It is well known that when $Q^{2}\rightarrow 0$, the heavy diquark
is seen by the gluon as a point particle without the inner
structure, and hence the form factor for the effective vertex should
be normalized to unity. When $Q^{2}\rightarrow \infty$, the gluon
can see the individual quarks inside the diquark, and hence the form
factor for the effective vertex should approach to zero. We
calculate the form factors with the BS wave functions obtained numerically for
the heavy diquarks. The dependence of the form factors
on the square of the momentum transfer is shown in Fig.~3. We can
see that the behavior of our results coincides with the tendency of
the above physical picture.

\begin{figure}[hbtp]\label{fig3}
\begin{center}
\includegraphics[scale=0.75]{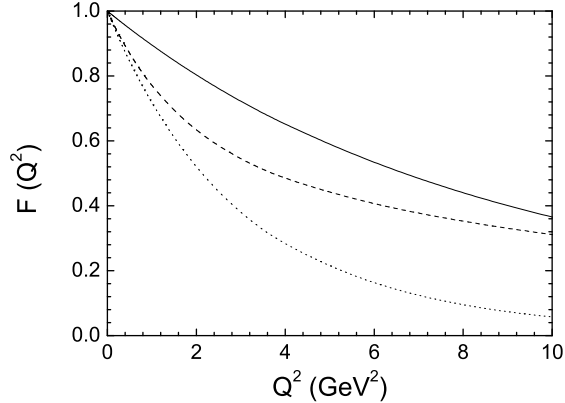}
\caption{The normalized form factors for the vertices of the heavy
diquark coupling to gluon as a function of the square of the
momentum transfer. The solid, dashed, and dotted lines represent the
heavy diquark composed of double $b$ quarks, $b$ and $c$ quarks, and
double $c$ quarks, respectively.}
\end{center}
\end{figure}

\section*{IV. BS equation for doubly heavy baryons}

As discussed in Sec. I, the doubly heavy baryon can be regarded as a
bound state of a heavy diquark and a light quark in the heavy quark
limit.

Let us define the ratios $\eta_{1}=m_{l}/(m_{l}+m_{D})$ and
$\eta_{2}=m_{D}/(m_{l}+m_{D})$ ($m_{l}$ is the light quark mass) for
convenience. The BS wave function for the doubly heavy baryon
composed of a scalar heavy diquark and a light quark is defined as
the following:
\begin{eqnarray}\label{eq:39}
  \chi_{P}(y_{1},y_{2})&=&\langle0|\psi_{l}(y_{1})\phi_{D}(y_{2})|P\rangle\nonumber\\
  &=&e^{-iPY}\int\frac{{\rm
  d}^{4}q}{(2\pi)^{4}}\chi_{P}(q)e^{-iqy},
\end{eqnarray}
where $\psi_{l}$ and $\phi_{D}$ stand for the light quark field and
the scalar heavy diquark field, respectively,
$Y\equiv\eta_{1}y_{1}+\eta_{2}y_{2}$ is the coordinate of the doubly
heavy baryon mass center, $y\equiv y_{1}-y_{2}$ is the relative
 coordinate, $P$ is the
momentum of the doubly heavy baryon, and $q$ is the relative
momentum between the heavy diquark and the light quark.

It is straightforward to derive the following BS equation for the
doubly heavy baryon containing a scalar heavy diquark and a light
quark:
\begin{eqnarray}\label{eq:40}
\chi_{P}(q)=S_{l}(q_{1})\int\frac{{\rm
d}^{4}q^{\prime}}{(2\pi)^{4}}G(P,q,q^{\prime})\chi_{P}(q^{\prime})S_{D}(q_{2}),
\end{eqnarray}
where $q_{1}=\eta_{1}P+q$ and $q_{2}=\eta_{2}P-q$ are the momenta of
the light quark and the heavy diquark, respectively, $S_{l}(q_{1})$
and $S_{D}(q_{2})$ are the propagators of the light quark and the
heavy diquark, respectively, $G$ is the kernel which is, motivated
by the potential model, given by~\cite{eic78}
\begin{eqnarray}\label{eq:41}
  -iG(P,q,q^{\prime})=I\otimes IV_{1}(q,q^{\prime})+\gamma_{\mu}\otimes
  \Gamma^{\mu}V_{2}(q,q^{\prime}),
\end{eqnarray}
where $\Gamma^{\mu}=(q_{2}^{\mu}+q_{2}^{\prime\mu})F(Q^{2})$ is the
effective vertex of a gluon with two scalar heavy diquarks, which
has been derived in Sec. III, $V_{1}$ and $V_{2}$ are the scalar
confinement and one-gluon-exchange terms given in the following
respectively (after imposing covariant instantaneous approximation
\cite{jin92,dai93,dai94,cha05}):
\begin{eqnarray}\label{eq:42}
   V_{1}(q_{t}-q^{\prime}_{t})=\frac{8\pi
  \kappa^{\prime}}{[-(q_{t}-q^{\prime}_{t})^{2}+\mu^{2}]^{2}}-(2\pi)^{3}\delta^{3}(q_{t}-q^{\prime}_{t})\int\frac{{\rm
d}^{3}k_{t}}{(2\pi)^{3}}\frac{8\pi
  \kappa^{\prime}}{[-(q_{t}-k_{t})^{2}+\mu^{2}]^{2}},
\end{eqnarray}
and
\begin{eqnarray}\label{eq:43}
   V_{2}(q_{t}-q^{\prime}_{t})=-\frac{16\pi}{3}\frac{\alpha_{s}}{(q_{t}-q^{\prime}_{t})^{2}-\mu^{2}},
\end{eqnarray}
where $q^{(\prime)}_{t}$is the transverse projection of the relative
momentum ($q$) along the baryon momentum ($P$), the second term of
$V_{1}$ is introduced to remove the infrared singularity near the
point $q_{t}=q_{t}^{\prime}$, and the small parameter $\mu$ is
introduced to avoid the divergence in the numerical calculations. As
discussed in Ref.~\cite{guo96}, the dimension of $\kappa^{\prime}$
is three and that of $\kappa$ in the meson case is two. This extra
dimension in $\kappa^{\prime}$ should be caused by nonperturbative
diagrams which include the form factor effects at low momentum
region. We expect that $\kappa^{\prime}\sim\Lambda_{QCD}\kappa$,
since $\Lambda_{QCD}$ is the only parameter related to confinement.
In the numerical calculations, we let $\kappa^{\prime}$ vary between
$0.01~GeV^{3}$ and $0.06~GeV^{3}$ \cite{guo99,guo07,guo0702}.

The light quark propagator can be written as the following form
\cite{guo08,cha05}:
\begin{eqnarray}\label{eq:44}
  S_{l}(q_{1})=i\rlap/v\Bigl[\frac{\Lambda_{l}^{+}}{\eta_{1}M+q_{l}-\omega_{l}+i\varepsilon}
  +\frac{\Lambda_{l}^{-}}{\eta_{1}M+q_{l}+\omega_{l}-i\varepsilon}\Bigr],
\end{eqnarray}
where $M$ and $v$ are respectively the mass and velocity of the
doubly heavy baryon, $q_{l}=q\cdot v$ and
$q_{t}^{\mu}=q^{\mu}-q_{l}v^{\mu}$ are the longitudinal and
transverse projections of the relative momentum (q) along the doubly
heavy baryon momentum (P), respectively,
$\omega_{l}=\sqrt{m_{l}^{2}-q_{t}^{2}}$, and $\Lambda_{l}^{\pm}$ are
the projection operators given by
\begin{eqnarray}\label{eq:45}
 \Lambda_{l}^{\pm}=
\frac{\omega_{l}\pm\rlap/v(\rlap/q_{t}+m_{l})}{2\omega_{l}}.
\end{eqnarray}

In the leading order of a $1/m_{Q}$ expansion, the propagator of the
scalar heavy diquark can be written as
\begin{eqnarray}\label{eq:46}
  S_{D}(q_{2})=\frac{i}{2m_{D}(\eta_{2}M-q_{l}-m_{D}+i\varepsilon)}.
\end{eqnarray}

After writing down the most general form for the BS wave function
and taking into account its property under parity transformation, we
can parametrize the BS wave function for the doubly heavy baryon
with a scalar heavy diquark and a light quark in the following form:
\begin{equation}\label{eq:47}
  \chi_{P}(q)=(g_{s_{1}}+{\rlap/q_{t}}g_{s_{2}})u(v),
\end{equation}
where $u(v)$ is the spinor of the doubly heavy baryon, $g_{s_{1}}$
and $g_{s_{2}}$ are the Lorentz-scalar functions of $q_{t}^{2}$,
$q_{l}$ and $P^{2}=M^{2}$.

Defining ${\tilde g_{s_{1(2)}}}(q_{t})\equiv\int\frac{{\rm d}
q_{l}}{2\pi}g_{s_{1(2)}}$, one finds that the BS scalar wave
functions satisfy the coupled integral equations as follows:
\begin{eqnarray}\label{eq:48}
  {\tilde g_{s_{1}}}(q_{t})&=&-\int\frac{{\rm
  d}^{3}q^{\prime}_{t}}{(2\pi)^{3}}\frac{(m_{l}+\omega_{l})
  [V_{1}(q_{t}-q^{\prime}_{t})+2m_{D}F(Q^{2})V_{2}(q_{t}-q^{\prime}_{t})]}
  {4\omega_{l}m_{D}(M-m_{D}-\omega_{l})}~{\tilde
  g_{s_{1}}}(q^{\prime}_{t})\nonumber\\
  &&-\int\frac{{\rm
  d}^{3}q^{\prime}_{t}}{(2\pi)^{3}}\frac{V_{1}(q_{t}-q^{\prime}_{t})-2m_{D}F(Q^{2})V_{2}(q_{t}-q^{\prime}_{t})}
  {4\omega_{l}m_{D}(M-m_{D}-\omega_{l})}~q_{t}\cdot q^{\prime}_{t}~{\tilde
  g_{s_{2}}}(q^{\prime}_{t}),
  \end{eqnarray}
  \begin{eqnarray}\label{eq:49}
  {\tilde g_{s_{2}}}(q_{t})&=&-\int\frac{{\rm
  d}^{3}q^{\prime}_{t}}{(2\pi)^{3}}\frac{
  V_{1}(q_{t}-q^{\prime}_{t})+2m_{D}F(Q^{2})V_{2}(q_{t}-q^{\prime}_{t})}
  {4\omega_{l}m_{D}(M-m_{D}-\omega_{l})}~{\tilde
  g_{s_{1}}}(q^{\prime}_{t})\nonumber\\
  &&-\int\frac{{\rm
  d}^{3}q^{\prime}_{t}}{(2\pi)^{3}}\frac{(m_{l}-\omega_{l})
  [V_{1}(q_{t}-q^{\prime}_{t})-2m_{D}F(Q^{2})V_{2}(q_{t}-q^{\prime}_{t})]}
  {4\omega_{l}m_{D}(M-m_{D}-\omega_{l})}\frac{q_{t}\cdot q^{\prime}_{t}}{q_{t}^{2}}~{\tilde
  g_{s_{2}}}(q^{\prime}_{t}).
\end{eqnarray}

The normalization condition for the doubly heavy baryon with a
scalar heavy diquark and a light quark is given by (after imposing
the covariant instantaneous approximation on the kernel)
\begin{eqnarray}\label{eq:50}
  i\delta^{i_{1}i_{2}}_{j_{1}j_{2}}\int\frac{{\rm d}^{4}q~{\rm d}^{4}q^{\prime}}{(2\pi)^{8}}{\bar
  \chi}_{P}(q,s)\Bigl[\frac{\partial}{\partial
  P_{0}}I_{P}(q,q^{\prime})\Bigr]^{i_{1}i_{2}j_{2}j_{1}}\chi_{P}(q^{\prime},s^{\prime})=\delta_{ss^{\prime}},
\end{eqnarray}
where $i_{1(2)}$ and $j_{1(2)}$ represent the color indices of the
heavy diquark and the light quark, respectively, $s^{(\prime)}$ is
the spin index for the doubly heavy baryon and
$I_{P}^{i_{1}i_{2}j_{2}j_{1}}$ is the inverse of the four point
propagator defined as follows:
\begin{eqnarray}\label{eq:51}
  I^{i_{1}i_{2}j_{2}j_{1}}_{P}(q,q^{\prime})=\delta^{i_{1}j_{1}}\delta^{i_{2}j_{2}}(2\pi)^{4}\delta^{4}(q-q^{\prime})S^{-1}_{l}(q_{1})S^{-1}_{D}(q_{2}).
\end{eqnarray}

After some algebra, Eq. (\ref{eq:50}) can be written in the
following form:
\begin{eqnarray}\label{eq:52}
  -\frac{i}{6}\int\frac{{\rm d}^{4}q}{(2\pi)^{4}}\Bigl\{{\rm Tr}
  [{\tilde \chi}_{P}(q_{t}){\tilde {\bar \chi}}_{P}(q_{t})S_{l}(q_{1})(-i\eta_{1}
  \rlap/\varepsilon)S_{l}(q_{1})S_{D}(q_{2})]\nonumber\\
  +{\rm Tr}[{\tilde \chi}_{P}(q_{t}){\tilde {\bar \chi}}_{P}(q_{t})(-2i\eta_{2})
  q_{2}\cdot\varepsilon S_{l}(q_{1})S^{2}_{D}(q_{2})]\Bigr\}=1,
\end{eqnarray}
where $\varepsilon=(1,{\vec 0})$, ${\tilde
  \chi}_{P}(q_{t})$ and ${\tilde{\bar
  \chi}}_{P}(q_{t})$ are the transverse projections of the BS
wave functions given as follows:
\begin{eqnarray}\label{eq:53}
  {\tilde
  \chi}_{P}(q_{t})=-iS_{l}(q_{1})^{-1}\chi_{P}(q)S_{D}(q_{2})^{-1},
\end{eqnarray}
and
\begin{eqnarray}\label{eq:54}
  {\tilde{\bar\chi}}_{P}(q_{t})=-iS_{D}(q_{2})^{-1}{\bar\chi}_{P}(q)S_{l}(q_{1})^{-1},
\end{eqnarray}
respectively.

Then, one can derive the transverse projections of the BS wave
functions from Eqs. (\ref{eq:53}) (\ref{eq:54}), respectively:
\begin{eqnarray}\label{eq:55}
 {\tilde
  \chi}_{P}(q_{t})=[{\tilde h}_{s_{1}}(q_{t})+{\rlap/q_{t}}{\tilde h}_{s_{2}}(q_{t})]u(v),
\end{eqnarray}
and
\begin{eqnarray}\label{eq:56}
 {\tilde{\bar
  \chi}}_{P}(q_{t})={\bar u}(v)[{\tilde h}_{s_{1}}(q_{t})+{\rlap/q_{t}}{\tilde h}_{s_{2}}(q_{t})],
\end{eqnarray}
where
\begin{eqnarray}\label{eq:57}
  {\tilde h}_{s_{1}}(q_{t})&=&\int\frac{{\rm d}^{3}q^{\prime}_{t}}{(2\pi)^{3}}
  [V_{1}(q_{t}-q_{t}^{\prime})+2m_{D}F(Q^{2})V_{2}(q_{t}-q^{\prime}_{t})]
  {\tilde g}_{s_{1}}(q^{\prime}_{t}),
\end{eqnarray}
and
\begin{eqnarray}\label{eq:58}
  {\tilde h}_{s_{2}}(q_{t})&=&\int\frac{{\rm d}^{3}q^{\prime}_{t}}{(2\pi)^{3}}
  [V_{1}(q_{t}-q_{t}^{\prime})-2m_{D}F(Q^{2})V_{2}(q_{t}-q^{\prime}_{t})]
  \frac{q_{t}\cdot q_{t}^{\prime}}{q_{t}^2}{\tilde g}_{s_{2}}(q^{\prime}_{t}).
\end{eqnarray}

After substituting Eqs.~(\ref{eq:55}) (\ref{eq:56}) into Eq.
(\ref{eq:52}) and integrating out the longitudinal momentum $q_{l}$,
the normalization condition can be written in the following form:
\begin{eqnarray}\label{eq:59}
  &&\int\frac{{\rm d}^{3}q_{t}}{(2\pi)^{3}}\frac{1}{24Mm_{D}\omega_{l}^{3}
  (-M+m_{D}+\omega_{l})^{2}}\Bigl\{
  (m_{l}+\omega_{l})(2\eta_{2}M\omega_{l}^{2}
  +E\eta_{1}(m_{l}m_{D}\nonumber\\
  &&-\omega_{l}m_{D}+2m_{l}\omega_{l}+M\omega_{l}-Mm_{l})
  +E\eta_{1}(-M+m_{D}+2\omega_{l})q_{t}^{2}){\tilde h}_{s_{1}}^{2}(q_{t})\nonumber\\
  &&+4\omega_{l}^{2}q_{t}^{2}(\eta_{2}M+\eta_{1}E){\tilde h}_{s_{1}}(q_{t})
  {\tilde h}_{s_{2}}(q_{t})\nonumber\\
  &&+(2\eta_{2}M(m_{l}-\omega_{l})q_{t}^{2}\omega_{l}^{2}
  +E\eta_{1}(-M+m_{D}+2\omega_{l})q_{t}^{4}\nonumber\\
  &&+E\eta_{1}(m_{l}-\omega_{l})(-m_{l}m_{D}-\omega_{l}m_{D}-2m_{l}\omega_{l}
  +M(m_{l}+\omega_{l}))q_{t}^{2}{\tilde h}_{s_{2}}^{2}(q_{t})\Bigr\}=1,
\end{eqnarray}
where $E=P\cdot \varepsilon$.

Now let us define the BS wave function for the doubly heavy baryon
composed of an axial-vector heavy diquark and a light quark as
follows:
\begin{eqnarray}\label{eq:60}
  \chi^{\mu}_{P}(y_{1},y_{2})&=&\langle0|\psi_{l}(y_{1})A^{\mu}_{D}(y_{2})|P\rangle\nonumber\\
  &=&e^{-iPY}\int\frac{{\rm
  d}^{4}q}{(2\pi)^{4}}\chi^{\mu}_{P}(q)e^{-iqy},
\end{eqnarray}
where $A^{\mu}_{D}(y_{2})$ stands for the axial-vector heavy diquark
field.

The BS equation for the doubly heavy baryon with an axial-vector
heavy diquark and a light quark is given by
\begin{eqnarray}\label{eq:61}
 \chi_{P}^{\mu}(q)=S_{l}(q_{1})\int\frac{{\rm
d}^{4}q^{\prime}}{(2\pi)^{4}}G_{\rho\nu}(P,q,q^{\prime})\chi_{P}^{\nu}(q^{\prime})S_{D}^{\mu\rho}(q_{2}),
\end{eqnarray}
where $S_{D}^{\mu\rho}(q_{2})$ is the propagator of the axial-vector
heavy diquark and $G_{\rho\nu}$ is the kernel for the BS equation
given by
\begin{eqnarray}\label{eq:62}
  -iG_{\rho\nu}(P,q,q^{\prime})=-g_{\rho\nu}I\otimes IV_{1}(q,q^{\prime})-\gamma^{\mu}\otimes
  \Gamma_{\mu\rho\nu}V_{2}(q,q^{\prime}),
\end{eqnarray}
where $\Gamma_{\mu\rho\nu}=(q_{2\mu}+q_{2\mu}^{\prime})g_{\rho\nu}
F(Q^{2})$ is the effective vertex for the axial-vector heavy diquark
coupling to the gluon, which was derived in Sec. III.

In the leading order of a $1/m_{Q}$ expansion, the propagator of the
axial-vector heavy diquark can be written as
\begin{eqnarray}\label{eq:63}
  S^{\mu\nu}_{D}(q_{2})=-i\frac{g^{\mu\nu}-v^{\mu}v^{\nu}}
  {2m_{D}(\eta_{2}M-q_{l}-m_{D}+i\varepsilon)}.
\end{eqnarray}

Similar to the case of the doubly heavy baryon containing a scalar
heavy diquark, we can parametrize the BS wave function for the
doubly heavy baryon containing an axial-vector heavy quark and a
light quark in the following form:
\begin{equation}\label{eq:64}
  \chi^{\mu}_{P}(q)=(g_{v1}+{\rlap/q_{t}}g_{v2})u^{\mu}(v),
\end{equation}
where $u^{\mu}(v)$ is the spinor of the heavy baryon, $g_{v1}$ and
$g_{v2}$ are the Lorentz-scalar functions of $q_{t}^{2}$, $q_{l}$
and $P^{2}=M^{2}$. When the spin of the doubly heavy baryon is
$1/2$,
$u^{\mu}(v)=\frac{1}{\sqrt{3}}(\gamma^{\mu}+v^{\mu})\gamma^{5}u(v)$,
while the spin of the doubly heavy baryon is $3/2$, $u^{\mu}(v)$ is
the Rarita-Schwinger vector spinor.

After defining ${\tilde g_{v_{1(2)}}}(q_{t})=\int\frac{{\rm
d}p_{l}}{2\pi}g_{v_{1(2)}}$, one can find that the BS scalar wave
functions satisfy the coupled integral equations in the following:
\begin{eqnarray}\label{eq:65}
  {\tilde g_{v_{1}}}(q_{t})&=&-\int\frac{{\rm
  d}^{3}q^{\prime}_{t}}{(2\pi)^{3}}\frac{(m_{l}+\omega_{l})
  [V_{1}(q_{t}-q^{\prime}_{t})+2m_{D}F(Q^{2})V_{2}(q_{t}-q^{\prime}_{t})]}
  {4\omega_{l}m_{D}(M-m_{D}-\omega_{l})}~{\tilde
  g_{v_{1}}}(q^{\prime}_{t})\nonumber\\
  &&-\int\frac{{\rm
  d}^{3}q^{\prime}_{t}}{(2\pi)^{3}}\frac{V_{1}
  (q_{t}-q^{\prime}_{t})-2m_{D}F(Q^{2})V_{2}(q_{t}-q^{\prime}_{t})}
  {4\omega_{l}m_{D}(M-m_{D}-\omega_{l})}~q_{t}\cdot q^{\prime}_{t}~{\tilde
  g_{v_{2}}}(q^{\prime}_{t}),
  \end{eqnarray}
  \begin{eqnarray}\label{eq:66}
  {\tilde g_{v_{2}}}(q_{t})&=&-\int\frac{{\rm
  d}^{3}q^{\prime}_{t}}{(2\pi)^{3}}\frac{
  V_{1}(q_{t}-q^{\prime}_{t})+2m_{D}F(Q^{2})V_{2}(q_{t}-q^{\prime}_{t})}
  {4\omega_{l}m_{D}(M-m_{D}-\omega_{l})}~{\tilde
  g_{v_{1}}}(q^{\prime}_{t})\nonumber\\
  &&-\int\frac{{\rm
  d}^{3}q^{\prime}_{t}}{(2\pi)^{3}}\frac{(m_{l}-\omega_{l})
  [V_{1}(q_{t}-q^{\prime}_{t})-2m_{D}F(Q^{2})V_{2}(q_{t}-q^{\prime}_{t})]}
  {4\omega_{l}m_{D}(M-m_{D}-\omega_{l})}\frac{q_{t}\cdot q^{\prime}_{t}}{q_{t}^{2}}~{\tilde
  g_{v_{2}}}(q^{\prime}_{t}).
\end{eqnarray}

The normalization condition for the BS wave function for the doubly
heavy baryon with an axial-vector heavy diquark and a light quark is
given by
\begin{eqnarray}\label{eq:67}
  i\delta^{i_{1}i_{2}}_{j_{1}j_{2}}\int\frac{{\rm d}^{4}q~{\rm d}^{4}q^{\prime}}{(2\pi)^{8}}{\bar
  \chi}^{\mu}_{P}(q,s)\Bigl[\frac{\partial}{\partial
  P_{0}}I_{P\mu\nu}(q,q^{\prime})\Bigr]^{i_{1}i_{2}j_{2}j_{1}}\chi^{\nu}_{P}
  (q^{\prime},s^{\prime})=\delta_{ss^{\prime}},
\end{eqnarray}
where $I^{i_{1}i_{2}j_{2}j_{1}}_{P\mu\nu}$ is the inverse of the
four point propagator
 defined as follows:
\begin{eqnarray}\label{eq:68}
  I_{P\mu\nu}^{i_{1}i_{2}j_{2}j_{1}}(q,q^{\prime})=\delta^{i_{1}j_{1}}\delta^{i_{2}j_{2}}(2\pi)^{4}\delta^{4}(q-q^{\prime})S^{-1}_{l}(q_{1})S^{-1}_{D\mu\nu}(q_{2}).
\end{eqnarray}

After some algebra, Eq. (\ref{eq:67}) can be written in the
following form:
\begin{eqnarray}\label{eq:69}
  &&-\frac{i}{6}\int\frac{{\rm d}^{4}q}{(2\pi)^{4}}\Bigl\{{\rm Tr}
  [{\tilde \chi}_{P\sigma}(q_{t}){\tilde {\bar \chi}}_{P\rho}
  (q_{t})S_{l}(q_{1})(-i\eta_{1}
  \rlap/\varepsilon)S_{l}(q_{1})S_{D}^{\sigma\rho}(q_{2})]\nonumber\\
  &&-{\rm Tr}[{\tilde \chi}_{P\sigma}(q_{t}){\tilde {\bar
  \chi}}_{P\rho}(q_{t})(i\eta_{2})
  (2q_{2}\cdot\varepsilon g_{\mu\nu}-\varepsilon_{\mu}q_{2\nu}-q_{2\mu}\varepsilon_{\nu})
   S_{l}(q_{1})S^{\sigma\nu}_{D}(q_{2})S^{\mu\rho}_{D}(q_{2})]\Bigr\}=1,
\end{eqnarray}
where ${\tilde
  \chi}_{P\sigma}(q_{t})$ and ${\tilde{\bar
  \chi}}_{P\rho}(q_{t})$ are the transverse projections of the BS
wave functions given by
\begin{eqnarray}\label{eq:70}
  {\tilde
  \chi}_{P\sigma}(q_{t})=-iS_{l}(q_{1})^{-1}\chi^{\alpha}_{P}(q)S_{D\alpha\sigma}(q_{2})^{-1},
\end{eqnarray}
and
\begin{eqnarray}\label{eq:71}
  {\tilde{\bar\chi}}_{P\rho}(q_{t})=-iS_{D\rho\beta}(q_{2})^{-1}{\bar\chi}^{\beta}_{P}(q)S_{l}(q_{1})^{-1},
\end{eqnarray}
respectively.

From Eqs.~(\ref{eq:70})(\ref{eq:71}) we get the expressions for the
transverse projections of the BS wave functions as follows:
\begin{eqnarray}\label{eq:72}
 {\tilde
  \chi}_{P\sigma}(q_{t})=[{\tilde h}_{v_{1}}(q_{t})+{\rlap/q_{t}}{\tilde h}_{v_{2}}(q_{t})]u_{\sigma}(v),
\end{eqnarray}
and
\begin{eqnarray}\label{eq:73}
 {\tilde{\bar
  \chi}}_{P\rho}(q_{t})={\bar u}_{\rho}(v)[{\tilde h}_{v_{1}}(q_{t})+{\rlap/q_{t}}{\tilde h}_{v_{2}}(q_{t})],
\end{eqnarray}
where
\begin{eqnarray}\label{eq:74}
  {\tilde h}_{v_{1}}(q_{t})&=&\int\frac{{\rm d}^{3}q^{\prime}_{t}}{(2\pi)^{3}}
  [V_{1}(q_{t}-q^{\prime}_{t})+2m_{D}F(Q^{2})V_{2}(q_{t}-q^{\prime}_{t})]{\tilde g}_{v_{1}}(q^{\prime}_{t}),
\end{eqnarray}
and
\begin{eqnarray}\label{eq:75}
  {\tilde h}_{v_{2}}(q_{t})&=&\int\frac{{\rm d}^{3}q^{\prime}_{t}}{(2\pi)^{3}}
  [V_{1}(q_{t}-q^{\prime}_{t})-2m_{D}F(Q^{2})V_{2}(q_{t}-q^{\prime}_{t})]
  \frac{q_{t}\cdot q^{\prime}_{t}}{q_{t}^{2}}{\tilde g}_{v_{2}}(q^{\prime}_{t}).
\end{eqnarray}

After substituting Eqs.~(\ref{eq:72})(\ref{eq:73}) into Eq.
(\ref{eq:69}) and integrating out the longitudinal momentum $q_{l}$,
the normalization condition can be written in the following form:
\begin{eqnarray}\label{eq:76}
  &&\int\frac{{\rm d}^{3}q_{t}}{(2\pi)^{3}}\frac{1}{24Mm_{D}\omega_{l}^{3}
  (-M+m_{D}+\omega_{l})^{2}}\Bigl\{
  (m_{l}+\omega_{l})[2\eta_{2}M\omega_{l}^{2}
  +E\eta_{1}(m_{l}m_{D}\nonumber\\
  &&-\omega_{l}m_{D}+2m_{l}\omega_{l}+M\omega_{l}-Mm_{l})
  +E\eta_{1}(-M+m_{D}+2\omega_{l})q_{t}^{2}]{\tilde h}_{v_{1}}^{2}(q_{t})\nonumber\\
  &&+4\omega_{l}^{2}q_{t}^{2}(\eta_{2}M+\eta_{1}E)
  {\tilde h}_{v_{1}}(q_{t}){\tilde h}_{v_{2}}(q_{t})\nonumber\\
  &&+(2\eta_{2}M(m_{l}-\omega_{l})q_{t}^{2}\omega_{l}^{2}
  +E\eta_{1}(-M+m_{D}+2\omega_{l})q_{t}^{4}\nonumber\\
  &&+E\eta_{1}(m_{l}-\omega_{l})(-m_{l}m_{D}-\omega_{l}m_{D}-2m_{l}\omega_{l}
  +M(m_{l}+\omega_{l}))q_{t}^{2}{\tilde
  h}_{v_{2}}^{2}(q_{t})\Bigr\}=1.
\end{eqnarray}

In our calculation, we take the constituent masses of the light
quarks as $m_{u}=m_{d}=0.33~GeV$, $m_{s}=0.45~GeV$, and the scale of
nonperturbative interaction $\Lambda_{QCD}\simeq0.2~GeV$. In order
to solve the coupled integral equations (\ref{eq:48}) (\ref{eq:49}),
we discretize the integration region into $n$ pieces (with $n$
sufficiently large). In this way the integral equations are
transformed into coupled matrix equations for the $n$ dimensional
vectors $\tilde{g}_{s_{1(2)}}$. Then, it is easy to obtain the
eigenvalue equations for $\tilde{g}_{s_{1(2)}}$. The same method is
applied in dealing with the coupled integral equations (\ref{eq:65})
(\ref{eq:66}). After solving the eigenvalue equations, we obtain
masses of the doubly heavy baryons shown in Table \ref{tab:2}. The
mass of the $\Xi_{cc}$ obtained in our model is consistent with the
experimental value, $3518.9\pm0.9~MeV$ \cite{ams08}. The obtained BS
scalar wave functions for the doubly heavy baryons composed of a
heavy diquark and a light quark are displayed in Fig. 4 and Fig. 5.
One finds that the masses and the amplitudes of BS scalar wave
functions for the doubly heavy baryons are independent of the spins
of both the heavy diquarks and the doubly heavy baryons. In fact,
this is just the consequence of the heavy diquark spin symmetry. So,
we redefine ${\tilde g}_{1}\equiv{\tilde g}_{s(v)_{1}}$ and ${\tilde
g}_{2}\equiv{\tilde g}_{s(v)_{2}}$ in Fig. 4 and Fig. 5 for
convenience.

\begin{table}[hbtp]
\caption{\label{tab:2} Values of the masses of baryons containing
two heavy quarks. The lower (upper) masses correspond to
$\kappa^{\prime}=0.01~(0.06)~GeV$.}
\begin{ruledtabular}
\begin{tabular}{ccccccc}
                     &$\Xi_{bb}$        &$\Xi_{bc}$       &$\Xi_{cc}$       &$\Omega_{bb}$       &$\Omega_{bc}$    &$\Omega_{cc}$\\
\hline
$M~(GeV)$            &10.08$\sim$10.10  &6.83$\sim$6.85   &3.52$\sim$3.56   &10.18$\sim$10.19    &6.94$\sim$6.95   &3.62$\sim$3.65\\
\end{tabular}
\end{ruledtabular}
\end{table}

\begin{figure}[hbtp]\label{fig4}
\begin{center}
\includegraphics[scale=1.8]{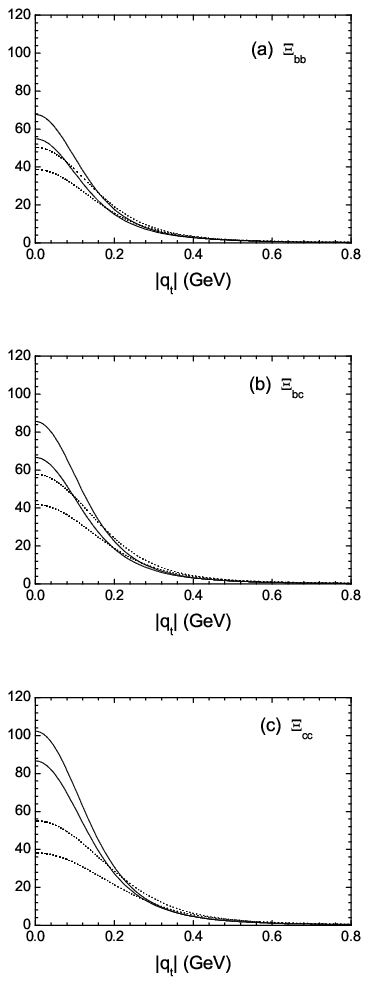}
\caption{The normalized BS scalar wave functions for the doubly
heavy baryons ($\Xi_{QQ^{\prime}}$) composed of a heavy diquark and
a light quark ($u,~d$). The solid (dotted) lines are for
$\kappa^{\prime}=0.01~(0.06)~GeV$. The upper solid and dotted lines
are for ${\tilde g}_{2}(q_{t})$ in unit of $GeV^{-1}$. The lower
solid and dotted lines are for ${\tilde g}_{1}(q_{t})$.}
\end{center}
\end{figure}

\begin{figure}[hbtp]\label{fig5}
\begin{center}
\includegraphics[scale=1.8]{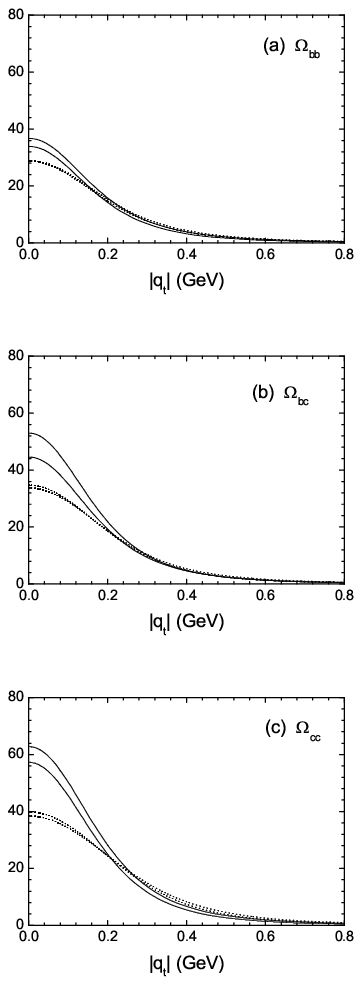}
\caption{The normalized BS scalar wave functions for the doubly
heavy baryons ($\Omega_{QQ^{\prime}}$) composed of a heavy diquark
and a strange light quark ($s$). The solid (dotted) lines are for
$\kappa^{\prime}=0.01~(0.06)~GeV$. The upper solid and dotted lines
are for ${\tilde g}_{2}(q_{t})$ in unit of $GeV^{-1}$. The lower
solid and dotted lines are for ${\tilde g}_{1}(q_{t})$.}
\end{center}
\end{figure}

\section*{V. Non-leptonic decay of doubly heavy baryons}

In this section, we will apply the obtained BS wave functions to
calculate the non-leptonic decay widths for the doubly heavy baryons
emitting a pseudo-scalar meson in the BS formalism. The Hamiltonian
describing such decays reads \cite{buc96}
\begin{eqnarray}\label{eq:77}
  H_{eff}=\frac{G_{F}}{\sqrt{2}}V_{cb}V_{UD}^{\ast}(a_{1}O_{1}+a_{2}O_{2}),
\end{eqnarray}
where $V_{cb}$ and $V^{\ast}_{UD}$ are the elements of
Cabibbo-Kobayashi-Maskawa matrix, $U$ and $D$ stand for the fields
of the light quarks involved in the decay, $O_{1}=[{\bar
D}\gamma^{\sigma}(1-\gamma^{5})U][{\bar
c}\gamma_{\sigma}(1-\gamma^{5})b]$ and $O_{2}=[{\bar
c}\gamma^{\sigma}(1-\gamma^{5})U][{\bar
D}\gamma_{\sigma}(1-\gamma^{5})b]$. $a_{1}$ and $a_{2}$ in Eq. (77)
are defined as the linear combination of Wilson coefficients
($c_{1}$ and $c_{2}$), $a_{1}=c_{1}+c_{2}/N_{c}$ and
$a_{2}=c_{2}+c_{1}/N_{c}$, where $N_{c}$ is an effective number of
colors which includes non-factorizable color-octet effects in the
hadronization process. Due to the lack of knowledge about
hadronization, $a_{1}$ and $a_{2}$ are treated as free parameters
and determined by fitting experimental data \cite{guo9802,lei03}.
 Since $b\rightarrow c$ decays are energetic,
the factorization assumption can be applied in our calculation.
Hence the decay amplitude of the two body non-leptonic decay becomes
the product of two matrix elements: one is related to the decay
constant of the factorized pseudo-scalar meson and the other is the
weak transition matrix between the initial and final doubly heavy
baryon states,
\begin{eqnarray}\label{eq:78}
  \langle\Xi_{Qc}^{(\ast)}(P_{f})P(k)|\Xi_{Qb}^{(\ast)}(P_{i})\rangle&=&
  -\frac{i}{\sqrt{2}}G_{F}V_{cb}V_{UD}^{\ast}a_{1}\int{\rm
  d}^{4}z\langle P(k)|{\bar D}(z)\gamma_{\sigma}
  (1-\gamma^{5})U(z)|0\rangle e^{-ikz}\nonumber\\
  &&\langle\Xi_{Qc}^{(\ast)}(P_{f})|{\bar
  b}(z)\gamma^{\sigma}(1-\gamma^{5})c(z)
  |\Xi_{Qb}^{(\ast)}(P_{i})\rangle,
\end{eqnarray}
where $P(k)$ stands for the pseudo-scalar meson with momentum $k$,
$\Xi_{QQ^{\prime}}^{(\ast)}$ stands for the doubly heavy baryon
composed of a (an) scalar (axial-vector) heavy diquark and a light
$u$ (or $d$) quark.

The first matrix element on the right hand side of Eq.~(\ref{eq:78})
is related to the decay constant of the pseudo-scalar meson,
$f_{P}$, which is defined as
\begin{eqnarray}\label{eq:79}
 \langle P(k)|{\bar
 D}(z)\gamma_{\sigma}(1-\gamma^{5})U(z)|0\rangle=-if_{P}k_{\sigma}.
\end{eqnarray}

The decay amplitude (Eq.~(\ref{eq:78})) is classified into three
cases according to different initial and final doubly heavy baryon
states as follows:
\begin{eqnarray}\label{eq:80-82}
 {\cal
 M}_{1}&=&\langle\Xi_{Qc}(P_{f})P(k)|\Xi_{Qb}^{\ast}(P_{i})\rangle,\\
{\cal
M}_{2}&=&\langle\Xi_{Qc}^{\ast}(P_{f})P(k)|\Xi_{Qb}^{\ast}(P_{i})\rangle,\\
{\cal
M}_{3}&=&\langle\Xi_{Qc}^{\ast}(P_{f})P(k)|\Xi_{Qb}(P_{i})\rangle.
\end{eqnarray}

\begin{figure}[hbtp]\label{fig6}
\begin{center}
\includegraphics[scale=0.8]{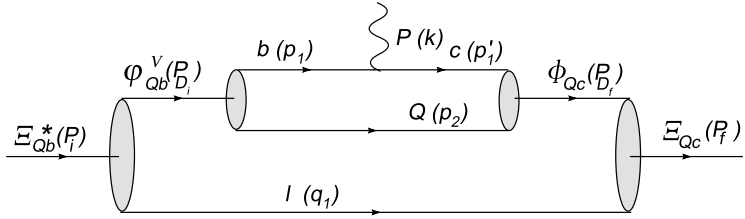}
\caption{The Feynman diagram for the non-leptonic decays of doubly
heavy baryons emitting a pseudo-scalar meson ($P$), taking the decay
amplitude ${\cal M}_{1}$ for instance. $
 \Xi_{Qc}(P_{f})$ and $\Xi^{\ast}_{Qb}(P_{i})$ stand for the states of
  the final and initial doubly heavy baryons with momenta
  $P_{f}$ and $P_{i}$, respectively.
  $\phi_{Qc}(P_{D_{f}})$ and $\varphi_{Qb}^{\nu}(P_{D_{i}})$
   stand for the diquark states
  involved in the final and initial doubly heavy baryons
  with momenta $P_{D_{f}}$ and $P_{D_{i}}$,
  respectively. $b(p_{1})$, $c(p_{1}^{\prime})$, $Q(p_{2})$, and $l(q_{1})$
  stand for different quark fields with corresponding momenta. $P(k)$ stands for
  the emitted meson with momentum $k$.}
\end{center}
\end{figure}

Let us first calculate the decay amplitude ${\cal M}_{1}$. The
Feynman diagram for ${\cal M}_{1}$ is shown in Fig. 6. ${\cal M}_{1}$
is related to the BS wave functions through the following equation:
\begin{eqnarray}\label{eq:83}
 {\cal
 M}_{1}&=&-\frac{1}{\sqrt{2}}G_{F}V_{cb}V_{UD}^{\ast}a_{1}f_{P}k_{\sigma}\int{\rm
 d}^{4}(x_{1}x_{2}y_{1}y_{2}z_{1}z_{2}z)e^{-ikz}\nonumber\\
 &&{\bar
 \chi}_{P_{f}}(x_{2},x_{1})S_{l}^{-1}(x_{1}-y_{1})
 \chi_{P_{i}}^{\mu}(y_{1},y_{2})S_{D}^{-1}(x_{2}-z_{1})S_{D\mu\nu}^{-1}(z_{2}-y_{2})\nonumber\\
&&\langle0|\phi_{Qc}(z_{1}){\bar
b}(z)\gamma^{\sigma}(1-\gamma^{5})c(z)\varphi_{Qb}^{\nu}(z_{2})|0\rangle,
\end{eqnarray}
where ${\bar
 \chi}_{P_{f}}$ and $\chi_{P_{i}}^{\mu}$ stand for the BS wave
 functions for the final and initial doubly heavy baryons, respectively,
  $\phi_{Qc}$ and $\varphi_{Qb}^{\nu}$ are the diquark field
  operators involved in the final and initial doubly heavy baryons,
  respectively.

The matrix element in Eq.~(\ref{eq:83}) is related to the BS wave
functions for the heavy diquarks through the following equation:
\begin{eqnarray}\label{eq:84}
 &&\langle0|\phi_{Qc}(z_{1}){\bar
b}(z)\gamma^{\sigma}(1-\gamma^{5})c(z)\varphi_{Qb}^{\nu}(z_{2})|0\rangle\nonumber\\
&=&\sum_{\lambda}\int\frac{{\rm d}^{4}P_{D_{i}}{\rm
d}^{4}P_{D_{f}}}{(2\pi)^{8}}(2\pi)^{2}\delta(P_{D_{f}l}-\sqrt{m_{D_{f}}^{2}-P_{D_{f}t}^{2}})
\delta(P_{D_{i}l}-\sqrt{m_{D_{i}}^{2}-P_{D_{i}t}^{2}})\nonumber\\
&&\times \frac{1}{6}\int\frac{{\rm d}^{4}p{\rm
d}^{4}p^{\prime}}{(2\pi)^{8}}(2\pi)^{4}\delta^{4}(p_{2}-p^{\prime}_{2}){\rm
Tr} [S(-p_{1}^{\prime}){\tilde {\tilde
\chi}}^{(c)}_{P_{D_{f}}}(p_{t}^{\prime})S(p_{2}){\tilde {\tilde
\chi}}^{(\lambda)}_{P_{D_{i}}}(p_{t})S(-p_{1})(1-\gamma^{5})\gamma^{\sigma}]\xi^{\nu(\lambda)}\nonumber\\
&&\times
e^{-iz(P_{D_{f}}-P_{D_{i}}+k)}e^{-iz_{1}P_{D_{f}}}e^{-iz_{2}P_{D_{i}}},
\end{eqnarray}
where ${\tilde {\tilde \chi}}^{(c)}_{P_{D_{f}}}$ and ${\tilde
{\tilde \chi}}^{(\lambda)}_{P_{D_{i}}}$ are the BS wave functions
for the heavy diquarks involved in the final and initial doubly
heavy baryons, respectively, $m_{D_{f}}$ and $m_{D_{i}}$ are the
masses of the heavy diquarks involved in the final and initial
doubly heavy baryons, respectively, $P_{D_{f(i)}l}=P_{D_{f(i)}}\cdot
v_{i}$ ($v_{i}$ denotes the velocity of the initial doubly heavy
baryon) and
$P_{D_{f(i)}t}^{\mu}=P_{D_{f(i)}}^{\mu}-P_{D_{f(i)}l}v_{i}^{\mu}$
are the longitudinal and transverse projections of the final
(initial) heavy diquark momentum ($P_{D_{f(i)}}$) along the initial
doubly heavy baryon momentum ($P_{i}$), respectively, and
$\xi^{\nu(\lambda)}$ is the $\lambda$-th polarization vector of the
axial-vector heavy diquark involved in the initial doubly heavy
baryon.

Substituting Eq.~(\ref{eq:84}) into Eq.~(\ref{eq:83}), the decay
amplitude ${\cal M}_{1}$ can be written in momentum space as
follows:

\begin{eqnarray}\label{eq:85}
  {\cal
  M}_{1}&=&-\frac{1}{\sqrt{2}}G_{F}V_{cb}V_{UD}^{\ast}a_{1}f_{P}k_{\sigma}
  (2\pi)^{4}\delta^{4}(P_{i}-P_{f}-k)\int\frac{{\rm d}^{4}q{\rm
d}^{4}q^{\prime}}{(2\pi)^{8}}(2\pi)^{4}\delta^{4}(q_{1}-q^{\prime}_{1})\nonumber\\
&&\times(2\pi)^{2}\delta(P_{D_{f}l}-\sqrt{m_{D_{f}}^{2}-P_{D_{f}t}^{2}})
\delta(P_{D_{i}l}-\sqrt{m_{D_{i}}^{2}-P_{D_{i}t}^{2}}) {\tilde {\bar
\chi}}_{P_{f}}(q^{\prime}_{t})S_{l}(q_{1}){\tilde
\chi}_{P_{i\nu}}(q_{t})\nonumber\\
&&\times\frac{1}{6}\sum_{\lambda}\int\frac{{\rm d}^{4}p{\rm
d}^{4}p^{\prime}}{(2\pi)^{8}}(2\pi)^{4}\delta^{4}(p_{2}-p^{\prime}_{2}){\rm
Tr} [S(-p_{1}^{\prime}){\tilde {\tilde
\chi}}^{(c)}_{P_{D_{f}}}(p_{t}^{\prime})S(p_{2}){\tilde {\tilde
\chi}}^{(\lambda)}_{P_{D_{i}}}(p_{t})S(-p_{1})(1-\gamma^{5})\gamma^{\sigma}]\xi^{\nu(\lambda)}.
\end{eqnarray}

Analogously, we can derive the other two decay amplitudes (${\cal
M}_{2}$ and ${\cal M}_{3}$) in the BS formalism as follows:
\begin{eqnarray}\label{eq:86}
  {\cal
  M}_{2}&=&-\frac{1}{\sqrt{2}}G_{F}V_{cb}V_{UD}^{\ast}a_{1}f_{P}k_{\sigma}
  (2\pi)^{4}\delta^{4}(P_{i}-P_{f}-k)\int\frac{{\rm d}^{4}q{\rm
d}^{4}q^{\prime}}{(2\pi)^{8}}(2\pi)^{4}\delta^{4}(q_{1}-q^{\prime}_{1})\nonumber\\
&&\times(2\pi)^{2}\delta(P_{D_{f}l}-\sqrt{m_{D_{f}}^{2}-P_{D_{f}t}^{2}})
\delta(P_{D_{i}l}-\sqrt{m_{D_{i}}^{2}-P_{D_{i}t}^{2}}){\tilde {\bar
\chi}}_{P_{f\tau}}(q^{\prime}_{t})S_{l}(q_{1}){\tilde
\chi}_{P_{i\nu}}(q_{t})\nonumber\\
&&\times\frac{1}{6}\sum_{\lambda\lambda^{\prime}}\int\frac{{\rm
d}^{4}p{\rm
d}^{4}p^{\prime}}{(2\pi)^{8}}(2\pi)^{4}\delta^{4}(p_{2}-p^{\prime}_{2}){\rm
Tr} [S(-p_{1}^{\prime}){\tilde {\tilde
\chi}}^{(\lambda^{\prime})(c)}_{P_{D_{f}}}(p_{t}^{\prime})S(p_{2}){\tilde
{\tilde
\chi}}^{(\lambda)}_{P_{D_{i}}}(p_{t})S(-p_{1})(1-\gamma^{5})\gamma^{\sigma}]
\xi^{\tau(\lambda^{\prime})}\xi^{\nu(\lambda)},\nonumber\\
\end{eqnarray}
and
\begin{eqnarray}\label{eq:87}
 {\cal
  M}_{3}&=&-\frac{1}{\sqrt{2}}G_{F}V_{cb}V_{UD}^{\ast}a_{1}f_{P}k_{\sigma}
  (2\pi)^{4}\delta^{4}(P_{i}-P_{f}-k)\int\frac{{\rm d}^{4}q{\rm
d}^{4}q^{\prime}}{(2\pi)^{8}}(2\pi)^{4}\delta^{4}(q_{1}-q^{\prime}_{1})\nonumber\\
&&\times(2\pi)^{2}\delta(P_{D_{f}l}-\sqrt{m_{D_{f}}^{2}-P_{D_{f}t}^{2}})
\delta(P_{D_{i}}-\sqrt{m_{D_{i}}^{2}-P_{D_{i}t}^{2}}) {\tilde {\bar
\chi}}_{P_{f\tau}}(q^{\prime}_{t})S_{l}(q_{1}){\tilde
\chi}_{P_{i}}(q_{t})\nonumber\\
&&\times\frac{1}{6}\sum_{\lambda^{\prime}}\int\frac{{\rm d}^{4}p{\rm
d}^{4}p^{\prime}}{(2\pi)^{8}}(2\pi)^{4}\delta^{4}(p_{2}-p^{\prime}_{2}){\rm
Tr} [S(-p_{1}^{\prime}){\tilde {\tilde
\chi}}^{(\lambda^{\prime})(c)}_{P_{D_{f}}}(p_{t}^{\prime})S(p_{2}){\tilde
{\tilde
\chi}}_{P_{D_{i}}}(p_{t})S(-p_{1})(1-\gamma^{5})\gamma^{\sigma}]\xi^{\tau(\lambda^{\prime})},
\end{eqnarray}
where $\xi^{\tau(\lambda^{\prime})}$ is the $\lambda^{\prime}$-th
polarization vector of the axial-vector heavy diquark involved in
the final doubly heavy baryon.

The differential decay width for the two body decay reads
\cite{ams08}
\begin{eqnarray}\label{eq:88}
 {\rm d}\Gamma=\frac{1}{32\pi^{2}}|{\cal M}|^{2}\frac{|{\bf
 k}|}{M_{i}^{2}}{\rm d}{\tilde\Omega},
\end{eqnarray}
where ${\tilde\Omega}$ denotes the solid angle, ${\cal M}$ stands
for the decay amplitude, $M_{i}$ is the mass of the initial baryon,
and $|{\bf k}|$ is the absolute value of the three momentum of the
particles in the final state in the rest frame of the initial state.

Numerically, the parameters in the Hamiltonian ($G_{F}$, $V_{cb}$,
and $V_{UD}$), and the masses and the decay constants of the
pseudo-scalar mesons are taken to have the following values
\cite{ams08}: $G_{F}=1.16637\times10^{-5}~GeV^{-2}$,
$V_{cb}=0.0412$, $V_{ud}=0.97418$, $V_{us}=V_{cd}=0.2255$,
$V_{cs}=0.9742$, $m_{\pi}=0.1396~GeV$, $m_{K}=0.4937~GeV$,
$m_{D}=1.8696~GeV$, $m_{D_{s}}=1.9685~GeV$, $f_{\pi}=0.1304~GeV$,
$f_{K}=0.1555~GeV$, $f_{D}=0.2058~GeV$, and $f_{D_{s}}=0.273~GeV$.
Our predictions for the non-leptonic decay widths for the doubly
heavy baryons emitting a pseudo-scalar meson are shown in Tables
\ref{tab:3}, \ref{tab:4}, \ref{tab:5}, and \ref{tab:6}, where the
superscripts $\frac{1}{2}$ and $\frac{3}{2}$ denote the spin of the
doubly heavy baryons.

\begin{table}[hbtp]
\caption{\label{tab:3} Predictions for the non-leptonic decay widths
for the doubly heavy baryons emitting $\pi$ meson.}
\begin{ruledtabular}
\begin{tabular}{cccc}
process                                                                       &decay width ($10^{-14}a_{1}^{2}GeV$)     &process                                                                      &decay width ($10^{-14}a_{1}^{2}GeV$)\\
\hline
$\Gamma(\Xi^{\ast\frac{1}{2}}_{bb}\rightarrow\Xi^{\frac{1}{2}}_{bc}\pi)$      &0.343$\sim$0.362                  &$\Gamma(\Omega^{\ast\frac{1}{2}}_{bb}\rightarrow\Omega^{\frac{1}{2}}_{bc}\pi)$      &0.591$\sim$0.607\\
$\Gamma(\Xi^{\ast\frac{1}{2}}_{bb}\rightarrow\Xi^{\ast\frac{1}{2}}_{bc}\pi)$  &0.205$\sim$0.211                  &$\Gamma(\Omega^{\ast\frac{1}{2}}_{bb}\rightarrow\Omega^{\ast\frac{1}{2}}_{bc}\pi)$  &0.380$\sim$0.381\\
$\Gamma(\Xi^{\ast\frac{3}{2}}_{bb}\rightarrow\Xi^{\ast\frac{3}{2}}_{bc}\pi)$  &4.110$\sim$4.234                  &$\Gamma(\Omega^{\ast\frac{3}{2}}_{bb}\rightarrow\Omega^{\ast\frac{3}{2}}_{bc}\pi)$  &7.606$\sim$7.643\\
$\Gamma(\Xi^{\frac{1}{2}}_{bc}\rightarrow\Xi^{\ast\frac{1}{2}}_{cc}\pi)$      &0.848$\sim$1.101                  &$\Gamma(\Omega^{\frac{1}{2}}_{bc}\rightarrow\Omega^{\ast\frac{1}{2}}_{cc}\pi)$      &1.708$\sim$1.876\\
$\Gamma(\Xi^{\ast\frac{1}{2}}_{bc}\rightarrow\Xi^{\ast\frac{1}{2}}_{cc}\pi)$  &0.415$\sim$0.587                  &$\Gamma(\Omega^{\ast\frac{1}{2}}_{bc}\rightarrow\Omega^{\ast\frac{1}{2}}_{cc}\pi)$  &0.965$\sim$1.019\\
$\Gamma(\Xi^{\ast\frac{3}{2}}_{bc}\rightarrow\Xi^{\ast\frac{3}{2}}_{cc}\pi)$  &8.626$\sim$12.110                 &$\Gamma(\Omega^{\ast\frac{3}{2}}_{bc}\rightarrow\Omega^{\ast\frac{3}{2}}_{cc}\pi)$  &19.435$\sim$20.529\\
\end{tabular}
\end{ruledtabular}
\end{table}

\begin{table}[hbtp]
\caption{\label{tab:4}Predictions for the non-leptonic decay widths
for the doubly heavy baryons emitting $K$ meson.}
\begin{ruledtabular}
\begin{tabular}{cccc}
process                                                                     &decay width ($10^{-15}a_{1}^{2}GeV$)     &process                                                                      &decay width ($10^{-15}a_{1}^{2}GeV$)\\
\hline
$\Gamma(\Xi^{\ast\frac{1}{2}}_{bb}\rightarrow\Xi^{\frac{1}{2}}_{bc}K)$      &0.265$\sim$0.267                  &$\Gamma(\Omega^{\ast\frac{1}{2}}_{bb}\rightarrow\Omega^{\frac{1}{2}}_{bc}K)$        &0.469$\sim$0.482\\
$\Gamma(\Xi^{\ast\frac{1}{2}}_{bb}\rightarrow\Xi^{\ast\frac{1}{2}}_{bc}K)$  &0.165$\sim$0.171                  &$\Gamma(\Omega^{\ast\frac{1}{2}}_{bb}\rightarrow\Omega^{\ast\frac{1}{2}}_{bc}K)$    &0.307$\sim$0.308\\
$\Gamma(\Xi^{\ast\frac{3}{2}}_{bb}\rightarrow\Xi^{\ast\frac{3}{2}}_{bc}K)$  &3.317$\sim$3.425                  &$\Gamma(\Omega^{\ast\frac{3}{2}}_{bb}\rightarrow\Omega^{\ast\frac{3}{2}}_{bc}K)$    &6.150$\sim$6.170\\
$\Gamma(\Xi^{\frac{1}{2}}_{bc}\rightarrow\Xi^{\ast\frac{1}{2}}_{cc}K)$      &0.649$\sim$0.845                  &$\Gamma(\Omega^{\frac{1}{2}}_{bc}\rightarrow\Omega^{\ast\frac{1}{2}}_{cc}K)$        &1.313$\sim$1.557\\
$\Gamma(\Xi^{\ast\frac{1}{2}}_{bc}\rightarrow\Xi^{\ast\frac{1}{2}}_{cc}K)$  &0.328$\sim$0.466                  &$\Gamma(\Omega^{\ast\frac{1}{2}}_{bc}\rightarrow\Omega^{\ast\frac{1}{2}}_{cc}K)$    &0.767$\sim$0.806\\
$\Gamma(\Xi^{\ast\frac{3}{2}}_{bc}\rightarrow\Xi^{\ast\frac{3}{2}}_{cc}K)$  &6.810$\sim$9.612                  &$\Gamma(\Omega^{\ast\frac{3}{2}}_{bc}\rightarrow\Omega^{\ast\frac{3}{2}}_{cc}K)$    &15.436$\sim$16.237\\
\end{tabular}
\end{ruledtabular}
\end{table}

\begin{table}[hbtp]
\caption{\label{tab:5}Predictions for the non-leptonic decay widths
for the doubly heavy baryons emitting $D$ meson.}
\begin{ruledtabular}
\begin{tabular}{cccc}
process                                                                     &decay width ($10^{-15}a_{1}^{2}GeV$)     &process                                                                    &decay width ($10^{-15}a_{1}^{2}GeV$)\\
\hline
$\Gamma(\Xi^{\ast\frac{1}{2}}_{bb}\rightarrow\Xi^{\frac{1}{2}}_{bc}D)$      &0.818$\sim$0.832                  &$\Gamma(\Omega^{\ast\frac{1}{2}}_{bb}\rightarrow\Omega^{\frac{1}{2}}_{bc}D)$        &1.404$\sim$1.466\\
$\Gamma(\Xi^{\ast\frac{1}{2}}_{bb}\rightarrow\Xi^{\ast\frac{1}{2}}_{bc}D)$  &0.693$\sim$0.741                  &$\Gamma(\Omega^{\ast\frac{1}{2}}_{bb}\rightarrow\Omega^{\ast\frac{1}{2}}_{bc}D)$    &1.292$\sim$1.319\\
$\Gamma(\Xi^{\ast\frac{3}{2}}_{bb}\rightarrow\Xi^{\ast\frac{3}{2}}_{bc}D)$  &13.885$\sim$14.834                  &$\Gamma(\Omega^{\ast\frac{3}{2}}_{bb}\rightarrow\Omega^{\ast\frac{3}{2}}_{bc}D)$  &25.873$\sim$26.419\\
$\Gamma(\Xi^{\frac{1}{2}}_{bc}\rightarrow\Xi^{\ast\frac{1}{2}}_{cc}D)$      &1.136$\sim$1.528                  &$\Gamma(\Omega^{\frac{1}{2}}_{bc}\rightarrow\Omega^{\ast\frac{1}{2}}_{cc}D)$        &2.434$\sim$2.552\\
$\Gamma(\Xi^{\ast\frac{1}{2}}_{bc}\rightarrow\Xi^{\ast\frac{1}{2}}_{cc}D)$  &0.945$\sim$1.464                  &$\Gamma(\Omega^{\ast\frac{1}{2}}_{bc}\rightarrow\Omega^{\ast\frac{1}{2}}_{cc}D)$    &2.383$\sim$2.426\\
$\Gamma(\Xi^{\ast\frac{3}{2}}_{bc}\rightarrow\Xi^{\ast\frac{3}{2}}_{cc}D)$  &19.525$\sim$30.028                 &$\Gamma(\Omega^{\ast\frac{3}{2}}_{bc}\rightarrow\Omega^{\ast\frac{3}{2}}_{cc}D)$   &47.895$\sim$48.741\\
\end{tabular}
\end{ruledtabular}
\end{table}

\begin{table}[hbtp]
\caption{\label{tab:6}Predictions for the non-leptonic decay widths
for the doubly heavy baryons emitting $D_{s}$ meson.}
\begin{ruledtabular}
\begin{tabular}{cccc}
process                                                                         &decay width ($10^{-13}a_{1}^{2}GeV$)     &process                                                                      &decay width ($10^{-13}a_{1}^{2}GeV$)\\
\hline
$\Gamma(\Xi^{\ast\frac{1}{2}}_{bb}\rightarrow\Xi^{\frac{1}{2}}_{bc}D_{s})$      &0.285$\sim$0.290                  &$\Gamma(\Omega^{\ast\frac{1}{2}}_{bb}\rightarrow\Omega^{\frac{1}{2}}_{bc}D_{s})$      &0.511$\sim$0.515\\
$\Gamma(\Xi^{\ast\frac{1}{2}}_{bb}\rightarrow\Xi^{\ast\frac{1}{2}}_{bc}D_{s})$  &0.253$\sim$0.271                  &$\Gamma(\Omega^{\ast\frac{1}{2}}_{bb}\rightarrow\Omega^{\ast\frac{1}{2}}_{bc}D_{s})$  &0.471$\sim$0.483\\
$\Gamma(\Xi^{\ast\frac{3}{2}}_{bb}\rightarrow\Xi^{\ast\frac{3}{2}}_{bc}D_{s})$  &5.065$\sim$5.435                &$\Gamma(\Omega^{\ast\frac{3}{2}}_{bb}\rightarrow\Omega^{\ast\frac{3}{2}}_{bc}D_{s})$    &9.437$\sim$9.668\\
$\Gamma(\Xi^{\frac{1}{2}}_{bc}\rightarrow\Xi^{\ast\frac{1}{2}}_{cc}D_{s})$      &0.368$\sim$0.497                  &$\Gamma(\Omega^{\frac{1}{2}}_{bc}\rightarrow\Omega^{\ast\frac{1}{2}}_{cc}D_{s})$      &0.794$\sim$0.828\\
$\Gamma(\Xi^{\ast\frac{1}{2}}_{bc}\rightarrow\Xi^{\ast\frac{1}{2}}_{cc}D_{s})$  &0.328$\sim$0.513                  &$\Gamma(\Omega^{\ast\frac{1}{2}}_{bc}\rightarrow\Omega^{\ast\frac{1}{2}}_{cc}D_{s})$  &0.828$\sim$0.851\\
$\Gamma(\Xi^{\ast\frac{3}{2}}_{bc}\rightarrow\Xi^{\ast\frac{3}{2}}_{cc}D_{s})$  &6.763$\sim$10.520                &$\Gamma(\Omega^{\ast\frac{3}{2}}_{bc}\rightarrow\Omega^{\ast\frac{3}{2}}_{cc}D_{s})$   &16.650$\sim$17.099\\
\end{tabular}
\end{ruledtabular}
\end{table}

\section*{VI. Summary and discussion}

In the heavy quark limit, a doubly heavy baryon can be regarded as a
bound state composed of a heavy diquark and a light quark. We first
establish the BS equations for both the heavy diquarks and the
doubly heavy baryons, respectively, in the leading order of a $1/m_{Q}$
expansion. The kernel for the BS equation contains the scalar
confinement and one-gluon-exchange terms, which are motivated by the
potential model and successfully used in the cases of mesons and
heavy baryons containing a single heavy quark. Since the size of the
heavy diquark is enhanced by ${\rm ln}^{2}m_{Q}$ with respect to
$1/m_{Q}$, we also introduce a few form factors to the effective
vertex for the heavy diquark coupling to the gluon in order to reflect
the inner structure of the heavy diquark.

The BS equations are solved numerically under the covariant
instantaneous approximation, which is suitable for the weakly bound
states of both the heavy diquark and the doubly heavy baryon. The
obtained masses of the doubly heavy baryons are consistent with
those from the lattice simulations \cite{lew01,mat0202}. It is found
that the properties of both the heavy diquarks and the doubly heavy
baryons are independent of their spin in the leading order of a
$1/m_{Q}$ expansion.

As we know, the superflavor symmetry relates doubly heavy baryons to
heavy mesons, and hence the form factors of the transitions of
doubly heavy baryons are reduced to the Isgur-Wise function, which
is well known for heavy mesons~\cite{san95,geo90,whi91}. The
calculation of the doubly heavy baryon transitions is greatly
simplified under the superflavor symmetry at the cost of ignoring
the derivation from non-pointlike spatial dispersion of the heavy
diquark. In this work, we directly calculate the decay amplitudes
for the doubly heavy baryons using the BS wave functions obtained
for both the heavy diquark and the doubly heavy baryons, instead of
employing the superflavor symmetry. We give the predictions for the
non-leptonic decay widths of doubly heavy baryons emitting a
pseudo-scalar meson. Our results will be tested in the future
experiments.

In our calculation, for the propagators of the heavy diquarks and
the light quarks involved in the bound states, we simply assume the
forms of free propagators with the masses of the heavy diquarks and
the light quarks taken to be the constituent ones. Actually, the
real propagators should be solved using the Dyson-Schwinger
equation. In such an approach, one has to guarantee the consistency
between the kernel of the BS equation and that of the
Dyson-Schwinger equation as required by the axial-vector
Ward-Takahashi identity \cite{rob07}. This is a very complicated
procedure and needs more careful investigations in the future.

At the HERA-B and Tevatron facilities more than $10^{5}$ events
involving double charm baryons are expected, while at the LHC one
can expect about $10^{9}$ events~\cite{ber98}. Since the available
energy at the LHC is much higher than the masses of $\Xi_{cc}$ and
$\Xi_{bb}$, it is believed that their production rates should be
comparable. The decay widths of the doubly heavy baryons we present
in this work will be tested in the forthcoming experiments.

Since the BS equations are established at the leading order in a
$1/m_{Q}$ expansion, we do not distinguish the different spins of both
the heavy diquarks and the doubly heavy baryons. Such differences
should happen at ${\cal O}(1/m_{Q})$. $1/m_{Q}$ corrections will be
studied in the future.

\begin{acknowledgments}
  One of the authors (M.-H. Weng) is very grateful to Dr. X.-H. Wu
  for a number of and very helpful discussions.
  This project is supported by the National Natural
  Science Foundation of China (Project Nos. 10675022 and 10975018) and
  the Special Grants from Beijing Normal University and by the Australian Research Council.
\end{acknowledgments}

\end{document}